\documentclass[fleqn]{2020SCGE}
\usepackage{graphicx}
\usepackage{comment}

\begin{document}
\ensubject{subject}

\newcommand{\bq}{\begin{equation}}
\newcommand{\eq}{\end{equation}}
\newcommand{\bqn}{\begin{eqnarray}}
\newcommand{\eqn}{\end{eqnarray}}
\newcommand{\nb}{\nonumber}
\newcommand{\lb}{\label}

\newcommand{\La}{\Lambda}
\newcommand{\va}{\scriptscriptstyle}
\newcommand{\be}{\nopagebreak[3]\begin{equation}}
\newcommand{\ee}{\end{equation}}

\newcommand{\ba}{\nopagebreak[3]\begin{eqnarray}}
\newcommand{\ea}{\end{eqnarray}}

\newcommand{\la}{\label}
\newcommand{\n}{\nonumber}
\newcommand{\su}{\mathfrak{su}}
\newcommand{\SU}{\mathrm{SU}}
\newcommand{\U}{\mathrm{U}}

\def\be{\nopagebreak[3]\begin{equation}}
\def\ee{\end{equation}}
\def\ba{\nopagebreak[3]\begin{eqnarray}}
\def\ea{\end{eqnarray}}
\newcommand{\f}{\frac}
\def\rmd{\rm d}
\def\lp{\ell_{\rm Pl}}
\def\d{{\rm d}}
\def\fe{\mathring{e}^{\,i}_a}
\def\fw{\mathring{\omega}^{\,a}_i}
\def\fq{\mathring{q}_{ab}}
\def\t{\tilde}

\def\db{\delta_b}
\def\dc{\delta_c}
\def\T{\mathcal{T}}
\def\GammaE{\Gamma_{\rm ext}}
\def\GammaEb{\bar\Gamma_{\rm ext}}
\def\GammaEh{\hat\Gamma_{\rm ext}}
\def\Hee{H_{\rm eff}^{\rm ext}}

\ArticleType{Article}
\Year{2024}
\Month{Jan}
\Vol{x}
\No{x}
\DOI{x}
\ArtNo{000000}

\title{
Nonexistence of quantum black and white hole horizons in an improved dynamic approach
}

\author[1,2,3,4]{Wen-Cong Gan}{{ganwencong@jxnu.edu.cn}}
\author[5]{Xiao-Mei Kuang}{{xmeikuang@yzu.edu.cn}}
\author[5]{Zhen-Hao Yang}{{yangzhenhao$_$yzu@163.com}}
\author[6]{Yungui Gong}{{gongyungui@nbu.edu.cn}}
\author[2]{Anzhong Wang}{{Anzhong$\_$Wang@baylor.edu; Corresponding author}}
\author[5,7]{Bin Wang}{{wang$_$b@sjtu.edu.cn}}

\AuthorMark{Gan W-C}


\AuthorCitation{Gan W-C}

\address[1]{College of Physics and Communication Electronics, Jiangxi Normal University, Nanchang 330022, China}
\address[2]{GCAP-CASPER, Physics Department, Baylor University, Waco, Texas 76798-7316, USA}
\address[3]{Institute for Theoretical Physics \& Cosmology, Zhejiang University of Technology, Hangzhou, 310023, China}
\address[4]{United Center for Gravitational Wave Physics (UCGWP),  Zhejiang University of Technology, Hangzhou, 310023, China}
\address[5]{Center for Gravitation and Cosmology, College of Physical Science and Technology, Yangzhou University, Yangzhou 225009, China}
\address[6]{Department of Physics, School of Physical Science and Technology, Ningbo University, Ningbo, Zhejiang 315211, China}
\address[7]{School of Aeronautics and Astronautics, Shanghai Jiao Tong University, Shanghai 200240, China}


\abstract{In this paper, we study the quantum geometric effects near the locations where classical black hole horizons used to appear in Einstein's classical theory, within the framework of an improved dynamic approach, in which the internal region of a black hole is modeled by the Kantowski-Sachs (KS) spacetime and the two polymerization parameters are functions of the phase space variables. Our detailed analysis shows that the effects are so strong that black and white hole horizons of the effective quantum theory do not exist at all and instead are replaced by transition surfaces, across which the metric coefficients and their inverses are smooth and remain finite, as are the corresponding curvatures, including the Kretschmann scalar.  These surfaces always separate trapped regions from anti-trapped regions. The number of such surfaces is infinite, so the corresponding KS spacetimes become geodesically complete, and no black and white hole-like structures exist in this scheme.}



\maketitle


\begin{multicols}{2}

\section{
Introduction
}
\label{sec:Intro}
\renewcommand{\theequation}{1.\arabic{equation}}\setcounter{equation}{0}

General relativity (GR) predicts its failure at spacetime singularities \cite{Hawking:1973uf}. Two well-known forms of singularities are the big-bang singularity in cosmology and the internal singularities of black holes. Quantum theories of gravity are expected to resolve these classical singularities by incorporating gravitational quantum effects; thus, the spacetimes near these singularities are still predictable quantum mechanically.

Loop quantum gravity (LQG) \cite{Ashtekar:2004eh,Thiemann:2007pyv} is one of the promising candidates of quantum gravity based on Hamiltonian formalism and canonical quantization of the holonomies of the connection and the fluxes of the triads. A very successful application of LQG is loop quantum cosmology (LQC), which was first considered in \cite{Bojowald:2001xe,Bojowald:2002gz,Ashtekar:2003hd} and then completed in \cite{Ashtekar:2006wn}. LQC is constructed by applying LQG techniques to cosmological models within the superminispace approach \cite{Ashtekar:2003hd}, and the resulting quantum corrections to classical geometry can be effectively described by a semiclassical effective Hamiltonian that incorporates the leading-order quantum geometric effects \cite{Taveras:2008ke}. The effective model works very well compared with the full quantum dynamics of LQC, even in the deep quantum regime \cite{Ashtekar:2011ni,Li:2023dwy,Agullo:2023rqq}, especially for states that sharply peaked on a classical trajectory at late times \cite{Kaminski:2019qjn}.
LQC resolves the big-bang singularity because of the fundamental result of LQG:  {\em quantum gravity effects always lead  the area operator to have a non-zero minimal area gap} \cite{Thiemann:2007pyv}. This non-zero area gap causes strong repulsive effects in the dynamics when the spacetime curvature reaches the Plank scale and the big-bang singularity is replaced by a quantum bounce \cite{Singh:2009mz}.

In LQC, two different quantization schemes exist, the so-called  $\mu_o$ and $\bar\mu$ schemes, which give different representations of quantum Hamiltonian constraints and lead to different effective dynamics \cite{Ashtekar:2011ni,Li:2023dwy,Agullo:2023rqq}. The fundamental difference between these two approaches is in the implementation of the minimal area gap mentioned above.
In the $\mu_o$ scheme, each holonomy $h_k^{(\mu)}$ is considered as an eigenstate of the area operator associated with the face of the elementary cell orthogonal to the $k$-th direction. The parameter $\mu$ is fixed by {requiring  the corresponding eigenvalue be the minimal area gap.} As a result, $\mu$ is a constant in this approach  \cite{Ashtekar:2003hd}.  However, it has been shown \cite{Corichi:2009pp} that this quantization does not have a proper semiclassical limit and suffers from dependence on the length $L_o$ of the fiducial cell. It also lacks consistent identified curvature scales. On the other hand, in the $\bar\mu$ scheme \cite{Ashtekar:2006wn}, the quantization of areas is referred to as the physical geometries, and when shrinking a loop until the minimal area enclosed by it, one should use the physical geometry. Since the latter depends on the phase space variables, when calculating the holonomy $h_k^{(\mu)}$, one finds that the parameter $\mu$ depends on the phase space variables, too. In the literature, this improved dynamical approach is often referred to as the $\bar\mu$ scheme, and has been shown to be the only scheme discovered so far that overcomes the limitations of the $\mu_o$ scheme and is consistent with observations \cite{Ashtekar:2011ni,Li:2023dwy,Agullo:2023rqq}.

In parallel to the studies of LQC, loop quantum black holes (LQBHs) have also been intensively studied in the past decade or so (See, for example, \cite{Olmedo:2016ddn,Ashtekar:2018cay,Ashtekar:2020ifw,Gambini:2022hxr,Ashtekar:2023cod} and references therein). In particular, since the Schwarzschild black hole interior becomes dynamical and the corresponding metric can be written in the form of the Kantowski-Sachs (KS) cosmological model [cf. Eq.(\ref{metric})], in which spacetime is homogeneous and the metric is only time-dependent, some LQC techniques can be borrowed to study black hole interiors directly. Along this line of thinking, LQBHs were initially studied within the  $\mu_o$ scheme \cite{Modesto:2004wm,Ashtekar:2005qt,Modesto:2005zm}. However, this LQBH model also suffers from limitations similar to those of the $\mu_o$ scheme in LQC \cite{Corichi:2015xia,Olmedo:2017lvt,Ashtekar:2018cay}. Soon the $\bar \mu$ scheme was applied to the Schwarzschild black hole interior by B\"ohmer and Vandersloot (BV) \cite{Boehmer:2007ket} (See also \cite{Chiou:2008eg,Chiou:2008nm} for a similar prescription in the KS universe, and \cite{Gambini:2013ooa,Gambini:2013hna,Gambini:2020nsf,Gambini:2020qhx,Liu:2021djf,Han:2022rsx} in the Painlev\'e-Gullstrand-like coordinates that cover both internal and external regions of the classical Schwarzschild black hole \footnote{For rigorous mathematical development of Ashtekar's formalism for spherically symmetric spacetimes and its loop quantization, see, for example, \cite{Bojowald:2004af,Bojowald:2004ag,Bojowald:2005cb,Campiglia:2007pr,Chiou:2012pg,Gambini:2020nsf} and references therein.}. Later, the $\bar \mu$ quantization scheme was shown to be the unique quantization scheme that is free from the dependence on the fiducial length $L_o$ and has consistent ultraviolet and infrared behavior \cite{Joe:2014tca}. It has universally bounded curvature scales and energy density, and the expansion and shear scalars are all finite, in addition to the geodesic completeness and generic resolution of strong singularities \cite{Saini:2016vgo}.

Despite these attractive features, the BV model suffers a severe drawback: {\em there are large departures from the classical theory very near the classical black hole horizon even for massive black holes, for which the curvatures at the horizon become very low} \cite{Boehmer:2007ket,Corichi:2015xia,Olmedo:2017lvt,Ashtekar:2018cay}. In addition, when the curvature reaches the Planck scale, the geometric radius of the round 2-spheres reaches a minimum and then bounces, giving rise to a transition surface ${\cal{T}}$, whereby the original singularity is replaced by a quantum bounce. The transition surface ${\cal{T}}$ naturally divides the spacetime into two regions, $T> T_{{\cal{T}}}$ and $T < T_{{\cal{T}}}$. To the past of ${\cal{T}}$, i.e., $T > T_{{\cal{T}}}$, we have a trapped region, and to its future ($T < T_{{\cal{T}}}$) an anti-trapped region appears, in which the geometric radius of the two spheres increases. However, in contrast to other LQBH models, this anti-trapped region is not bounded by a white-hole-like horizon; instead, it is followed by another bounce, across which the region becomes trapped again, and the radius of the two spheres starts to decrease \cite{Boehmer:2007ket,Chiou:2008eg,Chiou:2008nm}. This process will be repeated indefinitely, and after each bounce, the geometric radius of the two spheres will decrease. Thus, the area of the two spheres will soon become smaller than the minimal area gap, whereby the model becomes self-inconsistent \cite{Ashtekar:2018cay}.

In this paper, we focus on the past of ${\cal{T}}$,  as the spacetimes in the pre-transition phase were already studied in detail first in the vacuum case \cite{Boehmer:2007ket,Chiou:2008eg,Chiou:2008nm} and then in the case filled with matter \cite{Joe:2014tca} or a cosmological constant \cite{Dadhich:2015ora}.
In all of these cases, the spacetimes approach the classical ``charged" Nariai solutions as $T \rightarrow -\infty$, in which the radii of the 2-spheres become constants asymptotically, but with values much smaller than the Planck scale. Therefore, the validity of this asymptotic behavior  is questionable \cite{Ashtekar:2018cay}. In this paper, we shall leave this question aside and study in detail the spacetimes to the past of ${\cal{T}}$ by focusing on the quantum geometric effects near the location, $T = T^{\rm{GR}}_H = \ln(2m)$,  at which the classical black hole horizon would appear, and especially into the possible development of black hole or white hole horizons \cite{Hawking:1973uf,Ashtekar:2018cay,Wang:2003bt,Wang:2003xa,Gong:2007md,Hayward:1999ek}. To our surprise, we find that such a horizon is never developed and instead is replaced by an infinite number of transition surfaces, across which the metric coefficients and their inverses are smooth and remain finite, along with their corresponding curvatures, such as the Kretschmann scalar. Each of these surfaces always separates a trapped region from an anti-trapped region.
Thus, in the BV model, the quantum geometric effects are so strong that the black hole horizon, which would appear classically at $T^{\rm{GR}}_H$ now disappears, and the resultant KS model covers the entire spacetime, which is consistent with what was obtained previously \cite{Saini:2016vgo}.

{\em It must be noted that a geodesically complete/maximal spacetime does not imply that no black/white horizons exist in such a spacetime}. A concrete counterexample is the  flat Friedmann-Lemaitre-Robertson-Walker (FLRW) Universe,
\bq
\lb{eq0.2}
ds^2 = - dt^2 + a^2(t)\left(dr^2 + r^2d^2\Omega\right),
\eq where  $-\infty < t < \infty, \; 0 \le r < \infty, \;  0 \le \theta \le \pi, \; 0 \le \phi \le 2\pi$, and  $d\Omega^2 \equiv d\theta^2 + \sin^2\theta d\phi^2$. By properly choosing the matter fields (possibly with exotic matter) in our universe, we can always make it geodesically maximal \cite{Hawking:1973uf}. To determine whether a horizon is formed or not, let us first introduce two unity vectors, $u_{\mu} = \delta_{\mu}^t$ and $s_{\mu} = r a(t)\delta_{\mu}^r$, which are  time- and space-like, respectively. Then, it can be shown that the two null vectors, $l_{\mu}^{\pm} \equiv \left(u_{\mu} \pm s_{\mu}\right)/\sqrt{2}$, define the outgoing and ingoing light rays moving along the radial directions. The expansions of the two null geodesic congruences are given by
\cite{Ashtekar:2018cay}
\bqn
\lb{eq0.2a}
\Theta^{\pm} \equiv m^{\mu\nu}\nabla_{\mu}l^{\pm}_{\nu} = \frac{\sqrt{2}}{r a}\left(- r \dot a \pm1\right),
\eqn where $m_{\mu\nu} \equiv g_{\mu\nu} + u_{\mu}u_{\nu} - s_{\mu}s_{\nu}$, and $\dot a \equiv da/dt$. Clearly, in an expanding universe ($\dot a > 0$), we always have $\Theta^{-} < 0$, but
\bqn
\lb{eq0.2b}
\Theta^{+}   = \frac{\sqrt{2}}{r a}\left(1- r \dot a \right) = \begin{cases}
> 0, & r < {\dot a}^{-1},\cr
= 0, & r = {\dot a}^{-1},\cr
< 0, & r > {\dot a}^{-1}.\cr
\end{cases}
\eqn
Thus, on the surface $r = {\dot a}^{-1}$ a marginally trapped surface is formed, which separates {the {\em trapped} region ($\Theta^{\pm} < 0$) from the {\em untrapped} one  ($\Theta^{+} >0, \; \Theta^{-} < 0$).} Then, by definition, it represents an apparent horizon
\cite{Gong:2007md,Hayward:1999ek}.

On the other hand, the de Sitter spacetime is geodesically complete (in its maximal extension), but horizons indeed exist in such a spacetime. Geodesic completeness is only a necessary condition for avoiding the appearance of spacetime singularities, but not for the existence of black/white horizons.
Geodesic completeness and the existence of black/white horizons are two different conceptions, and their mathematical definitions are also different.
Yet, the existence of {\em a  transition surface}  is also irrelevant to whether spacetime is geodesically complete or not. Recall that {\em a transition surface is a surface that separates a trapped region ($\Theta^{\pm} < 0$)
from an anti-trapped one ($\Theta^{\pm} > 0$)} \cite{Ashtekar:2018cay} \footnote{ Anti-trapped regions are also called past-trapped regions in the literature \cite{Wang:2003bt,Wang:2003xa,Hayward:1999ek}.}.

It must also be noted that the range of the coordinate $t$ in the metric (\ref{eq0.2})  is determined by the requirement that {\em the spacetime must be geodesically complete or maximal}, that is, a spacetime in which a geodesic (timelike, spacelike or null) always ends either at infinities or at singularities \cite{Hawking:1973uf}. For example, for the radiation-dominated spacetime, we have $a(t) = a_0 (t/t_0)^{1/2}$, where $a_0$ and $t_0$ are two positive constants. Clearly, the spacetime becomes singular at $t = 0$, which causally separates the region $t > 0$ from the one $t < 0$ causally. Our expanding universe corresponds to the region $t > 0$, which, by definition, forms a geodesically maximal spacetime, as now $t = \infty$ forms the future boundaries, while the singularity at $t = 0$ forms the past boundary.
On the other hand, for the de Sitter spacetime, $a(t) = a_0e^{H(t - t_0)}$, the geodesically complete condition requires $t \in (-\infty, \infty)$. However, when $ t = -\infty$, we have $a(t = -\infty) = 0$, which indicates that some singularities still exist.
More detailed analyses show that this represents a cosmological horizon, and a further extension beyond $t = -\infty$ is still needed. In fact, the ($t, r$)-plane with $ -\infty < t < \infty$ and $0 \le r < \infty$  covers only half of the geodesically complete spacetime. For details, we refer readers to  \cite{Hawking:1973uf}.  However, these two examples clearly show how the range of the coordinate $t$ is determined for a given solution.

The same arguments can be equally applied to the KS spacetime described by the metric given by Eq.(\ref{metric}) below. In particular, for the Schwarzschild solution inside the black hole, the metric coefficients are given by Eq.(\ref{eq6d}), for which some become singular at both $T = -\infty$ and $T = T^{\rm{GR}}_H$.  Therefore, in this case $T$ is restricted to $T \in (-\infty, T^{\rm{GR}}_H)$. Further considerations of physical quantities, such as the Kretschmann scalar, show that  the  singularity at  $T = -\infty$ is a spacetime singularity that forms part of the boundaries of the geodesically maximal
Schwarzschild spacetime. On the other hand, the singularity at $T^{\rm{GR}}_H$ is only a coordinate singularity because no physical quantity becomes singular at this point, including the Kretschmann scalar. Therefore, to have a geodesically maximum spacetime, extensions of the spacetime across this surface are needed. The requirement that the extension across this surface be analytically unique determines that the extension must be the Kruskal-Szekeres extension \cite{Hawking:1973uf}.
On the other hand, for a given solution, if none of the singularities appear, the time-like coordinate $T$ should be taken as the whole range $T \in(-\infty, \infty)$. Of course, similar to the de Sitter spacetime written in the ($t, r$) introduced in Eq.(\ref{eq0.2}), this does not mean that the spacetime is already geodesically complete, and additional considerations are needed. In particular, for the BV solution to be considered in this paper, we find that the range $T \in(-\infty, \infty)$ corresponds to the range $p_c \in (\bar{p}_c, \infty)$, where
$\sqrt{|p_c|}$ represents the geometric radius of the 2-spheres of $T, x = $ Constant, and $\bar{p}_c$ is the minimal value of $p_c$ obtained asymptotically as $T \rightarrow - \infty$  [Eq.(\ref{CN})]. *Because $p_c(T = \infty) = \infty$, so  $T = \infty$ already represents the future boundaries of spacetime. Hence, the BV solution with $-\infty < T, \; x < \infty$ is geodesically complete, and the corresponding Penrose diagram is given in Fig. \ref{fig8} (b).

\begin{figure*}[htbp]
\begin{tabular}{cc}
\includegraphics[width=0.45\textwidth]{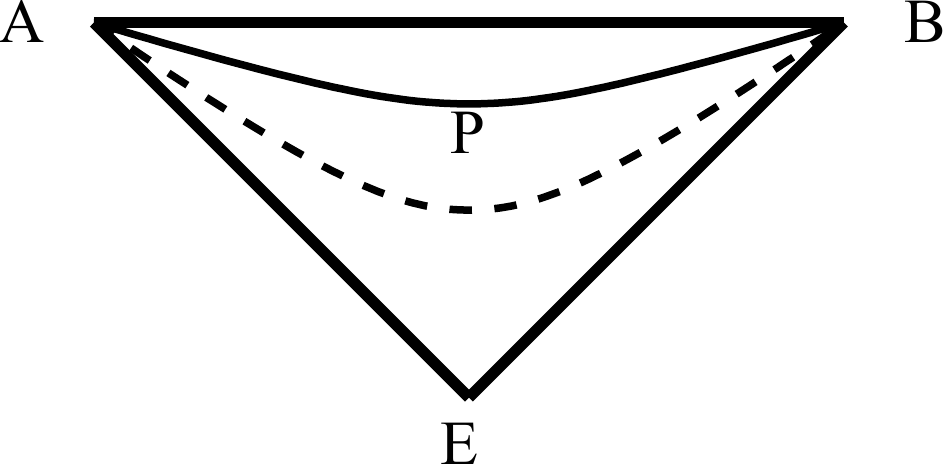}&
\includegraphics[width=0.45\textwidth]{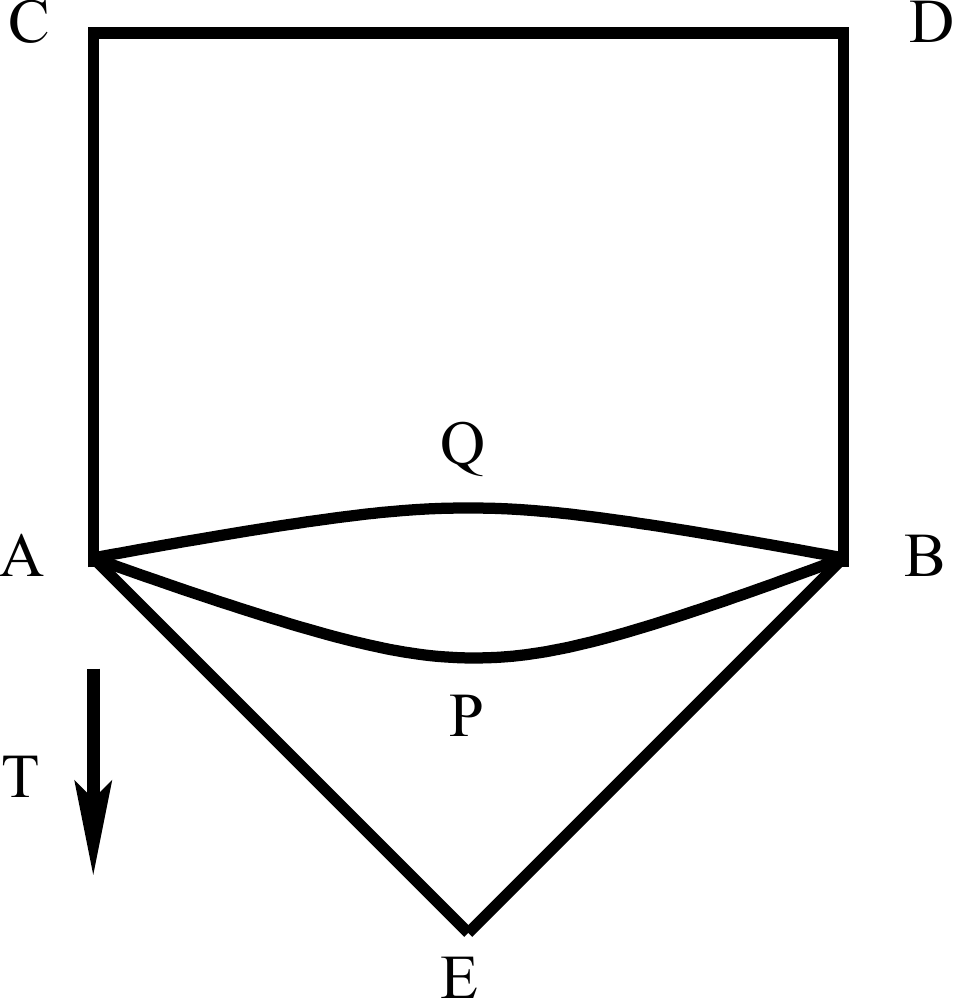}\\[6pt]
(a)&  (b) \\[6pt]

	\end{tabular}
\caption{(a) The Penrose diagram for the Kantowski-Sachs spacetime in classical Einstein's gravity. The horizontal line $AB$ represents the spacetime singularity. In the Schwarzschild case, it corresponds to   $p_c^{\rm{GR}}(T = -\infty) = 0$. The curve $APB$ corresponds to a $T=$ Constant surface with a nonzero radius.
(b) Penrose diagram for the BV model. Due to quantum geometric effects, the classical singularity that used to appear at $T = -\infty$ now is replaced by the transition surface  ${\cal{T}}$ denoted by the curve $APB$, at which we have $p_c(T_{\cal{T}}) > 0$. The quantum geometric effects are large in the region between the two curves $AQB$ and $APB$. The 2D plane with $\theta, \; \phi =$ Constant is asymptotically approaching a 2D de Sitter spacetime with a fixed radius $\sqrt{\bar{p}_c}$ as $T \rightarrow -\infty$.
In the region $T > T_{\cal{T}}$ the spaceitme is geodesically complete, and a black (or white) hole like horizon is never developed.}
\lb{fig8}
\end{figure*}

\vspace{.5cm}

With the above in mind, a natural question is whether a black/white hole horizon exists in the geodesically complete BV spacetime.
To answer this question, we study the existence of marginally trapped surfaces ($\Theta^{+} \Theta^{-} =  0$), a necessary condition for the existence of a  horizon or a transition surface  \cite{Hawking:1973uf,Ashtekar:2018cay,Wang:2003bt,Wang:2003xa,Gong:2007md,Hayward:1999ek}.
We find that such surfaces indeed exist. However, they represent neither black nor white hole horizons, but transition surfaces, as {\em they always separate trapped regions from anti-trapped ones or vice versa} \cite{Ashtekar:2018cay}, instead of separating trapped (anti-trapped) regions from un-trapped ones, as a black (white) hole horizon does \cite{Hawking:1973uf,Wang:2003bt,Wang:2003xa,Gong:2007md,Hayward:1999ek}. As a result, a black or white hole horizon does not exist. 
To the best of our knowledge, this is the first study in the literature to show this explicitly.

Specifically, the paper is organized as follows: In Sec. \ref{bi} we first briefly introduce the internal region of the classical Schwarzschild black hole, whereby we obtain the classical Hamiltonian with a special choice of the lapse function. An effective loop quantum Hamiltonian can then be obtained within the framework of the superminispace quantization scheme  \cite{Ashtekar:2003hd}, by replacing the two canonical phase space variables $b$ and $c$ via the relations
\bq
\lb{eq0.1}
b \rightarrow \frac{\sin(\delta_b b)}{\delta_b}, \quad c \rightarrow \frac{\sin(\delta_c c)}{\delta_c},
\eq in the classical lapse function and Hamiltonian, where
$\delta_b$ and $\delta_c$ are the two polymerization parameters, which are general functions of the phase space variables, that is,
\bq
\lb{eq0.1a}
\delta_b = \delta_b(b, p_b; c, p_c), \quad \delta_c = \delta_c(b, p_b; c, p_c).
\eq
In the case in which $\delta_b$ and $\delta_c$ are functions of $p_b$ and $p_c$ only,   the dynamical equations are given by Eqs.(\ref{eq2.25m}) - (\ref{eq2.25p})
(Note that the BV scheme belongs to the latter.).

In Sec. \ref{sec-bv}, we consider the BV model by first introducing the BV prescription of the two polymerization parameters $\delta_b$ and $\delta_c$ given by Eq.(\ref{eq3}), and then write down the corresponding dynamical equations, given explicitly by Eqs.(\ref{eq2.25s}) - (\ref{eq2.25v}). To estimate the region where the quantum effects near the classical black hole horizon become important, we first introduce a parameter $\epsilon$ via the relation $T_{\epsilon} = T^{\rm{GR}}_H(1 - \epsilon)$ at which $\left|\delta_c \right|_{T_{\epsilon}} \simeq {\cal{O}}(1)$, from which we find that such effects become important only very close to $T^{\rm{GR}}_H$ for massive black holes.
To study such effects explicitly, the choice of initial time and conditions is crucial. In this study, we choose the initial time $T_i$ that is far from both the transition surface and the classical black horizon,
$T_{\cal{T}} \ll T_i \ll T^{\rm{GR}}_H$, so that the initial conditions are as close to those of classical theory as possible [cf. Table \ref{ts2}]. With these initial conditions, we study the evolution of the dynamical equations and find that the metric coefficients remain finite and non-singular within any given finite time, even for initial conditions that are almost identical to their classical values.  These results are strongly supported by the analytical studies carried out in \cite{Saini:2016vgo}, in which it was shown explicitly that {\em  the BV spacetime is geodesically complete}. In such a spacetime, we further showed that there exist infinite number of transition surfaces that always separate trapped regions from untrapped ones, but no black hole or white hole horizons exist, as the latter always separates a trapped region from an untrapped one. This also explains why spacetime is now geodesically complete.

Finally, in Sec. \ref{conclusion} we present our main conclusions.

\begin{table*}[!ht]
\caption{Initial values of $c(T_i)$ calculated from Eq.(\ref{GRVsB}) at different times $T_i$ and the corresponding classical values  $c^{\rm{GR}}(T_i)$ for $m = M/G = \ell_{pl}$, for which we have   $T_{\cal{T}} \simeq -1.49$ and $T^{\rm{GR}}_H \simeq 0.693$.}
\centering
\begin{tabular}{|c|c|c|}
\hline
$T_i$ & $c(T_i)$ & $c^{\rm{GR}}(T_i)$ \\ \hline
1.45 & $~$ 2.68044 - 2.09266 I $~$ & -4.31636 \\ \hline
0.3 & -0.13272 & -0.130343 \\ \hline
$~0.643~$ & -0.0682837 & -0.0656195 \\ \hline
0.685 & $~$ -0.0883704 +  0.0213373 I  $~$ & -0.0603504 \\ \hline
\end{tabular}\lb{ts2}
\end{table*}

Before proceeding further, we would like to note that  in \cite{Ashtekar:2018cay} another interesting scheme  was mentioned, for which $\delta_b$ and $\delta_c$ were chosen as \cite{Corichi:2015xia}
\bqn
\lb{eq4.3}
\delta_b = \frac{\sqrt{\Delta}}{2m}, \quad L_o \delta_c = \sqrt{\Delta},
\eqn where $\Delta \; (\equiv  {4}\sqrt{3}\pi \gamma \ell_{pl}^2)$ denotes  the minimum area gap predicted by LQG, and $\ell_{pl}$ the Planck length. The parameter $\gamma$ denotes the Barbero-Immirzi parameter, which is fixed to $\gamma \simeq 0.2375$ by black hole entropy arguments \cite{Meissner:2004ju}.
As it was shown in \cite{Corichi:2015xia}, for such a choice, several desirable features exist, including that the corresponding physics is independent of $L_o$ and that a white/black hole structure exists. However, some undesirable aspects exist, such as the radii of the white and black hole horizons can be significantly different, and the curvature at the transition surface can be arbitrarily small for a massive black hole, i.e., the transition surface can emerge from a very low curvature regime. In \cite{Olmedo:2017lvt}, a generalization of
Eq.(\ref{eq4.3}) was further considered by replacing $(\delta_b, \delta_c) \rightarrow (\alpha\delta_b, \beta\delta_c)$ with $\alpha, \; \beta$ being dimensionless constants. In particular, they considered three particular cases: (i) $\beta = 1$; (ii) $\alpha = 1$; and (iii) $\alpha\beta =1$. However, none of them possess all of the desired features of the AOS model \cite{Ashtekar:2018cay}. For more details, please refer readers to \cite{Ongole:2023pbs},
in which the most general solutions of effective LQBHs with
$\delta_b,\; \delta_c$ being constant were studied systematically, and a large class of solutions that possess all the desired features of the AOS solution was identified.

In this paper, the Planck length $\ell_{pl}$ and mass $M_{pl}$ are defined, respectively, by $\ell_{pl} \equiv \sqrt{G\hbar/c^3}$ and $M_{pl} \equiv \sqrt{\hbar c/G}$, where $G$ denotes the Newtonian constant, $\hbar$ is the Planck constant divided by $2\pi$, and $c$ is the speed of light (Note that  in the main text, $c$ will be used to denote a phase space variable, and only in this paragraph we use it to denote the speed of light, without causing any confusion.).
Thus, in terms of the fundamental units, $M$, $L$, and $T$, the units of $\hbar$  and $c$ are $\left[\hbar\right] = M L^2T^{-1}, \; \left[c\right] = LT^{-1}$, where $M$, $L$, and $T$ denote the units of mass, length, and time, respectively. In this paper we shall adopt the natural units, so that  $\hbar = c  = 1$. Then, we find that $L = T, \; M = L^{-1}$, $\left[G\right]=L^3 M^{-1} T^{-2}=L^2$. In addition, all figures will be plotted in the units of $\ell_{pl}$ and $M_{pl}$, whenever length and mass are involved.

\section{
Internal Spacetimes of the Loop Quantum Black Holes
}
\lb{bi}
\renewcommand{\theequation}{2.\arabic{equation}}\setcounter{equation}{0}

The internal spacetime of a spherically symmetric black hole can always be written in the KS form \cite{Boehmer:2007ket}
\bq
\lb{metric}
ds^2 = - N^2 dT^2 + \frac{p_b^2}{L_o^2 \left|p_c\right|} dx^2 + \left|p_c\right| d\Omega^2,
\eq where $N, \; p_b, \; p_c$ are all functions of $T$ only,   and $L_o$ denotes the length of the fiducial cell in the $x$-direction.
The function $N$ is often called the lapse function, and $p_b$ and $p_c$ are dynamic variables that satisfy the Poisson brackets
\bqn
\lb{PBs}
\{c,p_c \}=2G \gamma,\quad  \{b,p_b \}=G \gamma,
\eqn where $b$ and $c$ are the corresponding phase space conjugate momenta of $p_b$ and $p_c$, respectively. The dimensions of the four phase space variables $\left(p_b, b; p_c, c\right)$ are
\bqn
\lb{DMsA}
\left[p_b\right] = \left[p_c\right] = L^2, \quad \left[b\right] = \left[c\right] = 1,
\eqn with $\left[x\right] = \left[L_o\right] = L$, where
``1" now means dimensionless.  The dimension of the coordinate $T$ depends on the choice of the gauge. In particular, for the gauge of Eq.(\ref{eq1}) given below, the lapse function has the dimension of length, while $T$ is dimensionless, i.e.
\bqn
\lb{DMsB}
\left[N^{\rm{GR}}\right] = L, \quad \left[T\right] = 1.
\eqn

Note that the KS metric (\ref{metric}) is invariant under the gauge transformations
\bq
\lb{GTs}
T = f(\hat T), \quad x = \alpha \hat x + x_o,
\eq via the redefinitions of the lapse function and the length of the fiducial cell,
\bq
\lb{RDs}
\hat N = Nf_{,\hat{T}}, \quad \hat{L}_o = \frac{L_o}{\alpha},
\eq where $f(\hat{T})$ is an arbitrary function of $\hat{T}$, and $\alpha$ and $x_o$ are arbitrary but real constants.
The equations of motion (EoMs) of the system can be obtained from the Hamiltonian equations
\bq
\lb{HEs}
\frac{d A}{dT} = \left\{A, H\right\},
\eq for any physical variable $A$ of the system.

\begin{table*}[ht!]
\caption{Different choices of the quantum parameters $\delta_b$ and $\delta_c$ for different quantization schemes.}
\centering
\begin{tabular}{|l|l|l|l|l|l|l|l|l|l|}
\hline
~ & $\mu_0$ & $\bar{\mu}'$ & CS & AOS & $\bar{\mu}$ & ABP & Modesto & BMM & GM \\ \hline
$\delta_b$ & $4\sqrt{3}$ & $\sqrt{\frac{\Delta}{p_b}}$ & $\frac{\sqrt{\Delta}}{r_o}$ & $\left( \frac{\sqrt{\Delta}}{\sqrt{2\pi}\gamma^2 m} \right)^{1/3}$ & $\sqrt{\frac{\Delta}{p_c}}$ & $\frac{\alpha}{R}=\frac{\alpha}{ \left|p_c\right|^{1/2}}$ & $\sigma(\delta)\delta$ & $f(O_b)$ & $f_b(O_b,O_c)$ \\ \hline
$\delta_c$ & $4\sqrt{3}$ & $\sqrt{\frac{\Delta}{p_c}}$ & $\frac{\sqrt{\Delta}}{L_o}$ & $\frac{1}{2 L_o} \left( \frac{\gamma \Delta^2}{4\pi^2 m} \right)^{1/3}$ & $\frac{\sqrt{\Delta {p_c}}}{p_b}$ & $\frac{\beta}{\Lambda}=\frac{\beta\left|p_c\right|^{1/2}}{p_b}$ & $\delta$ & $f(O_c)$ & $f_b(O_b,O_c)$ \\ \hline
~ & \cite{Ashtekar:2005qt} & \cite{Chiou:2008nm} & \cite{Corichi:2015xia} & \cite{Ashtekar:2018lag,Ashtekar:2018cay} & \cite{Boehmer:2007ket} & \cite{Alesci:2019pbs,Alesci:2020zfi,Gan:2022mle} & \cite{Modesto:2008im,Yan:2022fkr,Liu:2020ola,Zhu:2020tcf} & \cite{Bodendorfer:2019xbp,Ongole:2022rqi} & \cite{Garcia-Quismondo:2021xdc,Ongole:2022rqi} \\ \hline
\end{tabular}
\lb{scheme}
\end{table*}

\subsection{
Classical Model
}

To facilitate our discussions of LQBHs, let us first briefly review the classical black hole solution in Einstein's theory of general relativity (GR).  It is convenient to choose the lapse function as follows:
\bqn
\lb{eq1}
N^{\rm{GR}}= \frac{\gamma \; {\rm{sgn}}(p_c)\left|p_c\right|^{1/2}}{b},
\eqn which can be always realized by the gauge freedom of Eq.(\ref{GTs}) without loss of generality.  The corresponding Hamiltonian is given by \cite{Ashtekar:2018cay}
\bqn
\lb{eq2}
H^{\rm{GR}}[N^{\rm{GR}}] &\equiv& N^{\rm{GR}} \mathcal{H}^{\rm{GR}}\nb\\
&=& -\frac{1}{2G \gamma}\left(2c\;p_c+\left(b+\frac{\gamma^2}{b}\right)p_b\right).~~~
\eqn
Thus, from Eq.(\ref{HEs}) we find
\bqn
\lb{eq3a}
\dot b&=&\{b,H^{\rm{GR}}[N^{\rm{GR}}]\}=G \gamma \frac{\partial H^{\rm{GR}}}{\partial p_b}=-\frac{1}{2b}\left(b^2+\gamma^2\right),\\
\lb{eq3b}
\dot c&=&\{c,H^{\rm{GR}}[N^{\rm{GR}}]\}=2G \gamma \frac{\partial H^{\rm{GR}}}{\partial p_c}=-2c,\\
\lb{eq3c}
\dot{p}_b&=&\{p_b,H^{\rm{GR}}[N^{\rm{GR}}]\}=-G \gamma \frac{\partial H^{\rm{GR}}}{\partial b}=\frac{p_b}{2b^2}\left(b^2-\gamma^2\right), ~~~~~~~~\\
\lb{eq3d}
\dot{p}_c&=&\{p_c,H^{\rm{GR}}[N^{\rm{GR}}]\}=-2G \gamma \frac{\partial H^{\rm{GR}}}{\partial c}=2p_c,
\eqn where now an overdot denotes the derivative with respect to $T$. Then, the integration of Eqs.(\ref{eq3a}), (\ref{eq3b}) and (\ref{eq3d})
yield, respectively
\bqn
\lb{eq4a}
b^{\rm{GR}}(T) &=& \pm \gamma \sqrt{e^{T_o-T} -1},  \\
\lb{eq4b}
c^{\rm{GR}}(T)&=& c_oe^{-2T},\\
\lb{eq4c}
p^{\rm{GR}}_c(T)&=&p_c^o e^{2T},
\eqn where $T_o,\; c_o$ and $p_c^o$ are three integration constants with $T_o$ and $c_o$ being dimensionless and $p_c^o$ of dimensions $L^2$. To find $p_b$, we can first substitute the above solutions to Eq.(\ref{eq3c}) and then integrate it to find $p_b$, which will yield an additional integration constant, i.e., $p_b^o$. However, this constant is not arbitrary and must be chosen so that the classical Hamiltonian given by Eq.(\ref{eq2}) vanishes identically. A more straightforward way is to submit the above solutions into the classical Hamiltonian Eq.(\ref{eq2}) and find $p_b$ directly from this constraint, $H^{\rm{GR}}[N^{\rm{GR}}] = 0$. In doing so, we obtain
\bqn
\lb{eq4d}
p^{\rm{GR}}_b(T)= \mp \frac{2c_o p_c^o}{\gamma } e^{T-T_o} \sqrt{e^{T_o-T}-1}.
\eqn
Hence, we have
\bqn
\lb{eq5}
N^{\rm{GR}} = \pm \frac{\rm{sgn}(p_c)\left|p_c^o\right|^{1/2}}{\sqrt{e^{T_o-T} -1}}e^T,
\eqn and
\bqn
\lb{eq6}
ds^2 &=& \left|p_c^o\right| \Bigg(- \frac{e^{2T}}{e^{T_o-T}-1} dT^2 \nb\\
&& + \frac{4c_o^2 e^{-2T_o}}{ \gamma^2 L_o^2}(e^{T_o-T}-1)dx^2+e^{2T}d\Omega^2\Bigg). ~~~~
\eqn
Thus, without loss of generality, we can always assume $p_c^o > 0$. Then, setting
$r = \sqrt{p_c^o} e^T$ and rescaling $x$ by
\bq
\lb{rescalingx}
x \rightarrow t \equiv - \frac{2c_o \sqrt{p_c^o} e^{-T_o}}{\gamma L_o} x,
\eq the above metric takes the form of the classical Schwarzschild solution
\bqn
\lb{eq6a}
ds^2 = - \frac{1}{\frac{2m}{r}-1} dr^2 + \left( \frac{2m}{r}-1 \right) dt^2 + r^2 d\Omega^2, ~~~~~~
\eqn with $r < 2m $ and $m \equiv \sqrt{p_c^o} e^{T_o}/2$, which is related to the mass of the black hole via the relation $M = m/G$. Thus, the parameter $m$ has the dimension, $[m] = M^{-1} = L$ for the units adopted in this study.

From the above analysis, we can see that the rescaling (\ref{rescalingx}) simply allows us to set
\bqn
\lb{eq6b}
\frac{2c_o \sqrt{p_c^o} e^{-T_o}}{\gamma L_o} = - 1.
\eqn
On the other hand, using the gauge residual $T \rightarrow \hat T = T + C_o$, we can always set $p_c^o = 1$, where $C_o \equiv (1/2)\ln p_c^o$. Certainly, this rescaling will lead to $T - T_o = \hat T - \hat T_o$ and $m \equiv \sqrt{p_c^o} e^{T_o}/2 = e^{\hat T_o}/2$, where $\hat T_o \equiv T_o + C_o$. In addition, $L_o$ does not appear in the dynamical equations (\ref{eq3a}) - (\ref{eq3d}). Therefore, without loss of generality, we can always set $L_o = 1$. In summary, the constants $p_c^o, \; c_o$ and  $L_o$ can be chosen as follows:
\bqn
\lb{eq6c}
p_c^o = 1, \quad L_o = 1, \quad c_o = -\frac{\gamma L_o  e^{T_o}}{2 \sqrt{p_c^o}} = - \gamma m,
\eqn without affecting the physics of the spacetimes of the corresponding dynamical equations. Hence, we obtain
\bqn
\lb{eq6d}
b^{\rm{GR}}(T) &=& \pm \gamma \sqrt{2me^{-T} -1},  \nb\\
p^{\rm{GR}}_b(T)&=& \pm e^{T} \sqrt{2me^{-T}-1},\nb\\
c^{\rm{GR}}(T)&=& - \gamma m e^{-2T},\quad
p^{\rm{GR}}_c(T)= e^{2T}.
\eqn
In the rest of this paper, without loss of generality, we shall choose the ``+" signs for both $b^{\rm{GR}}(T)$ and $p^{\rm{GR}}_b(T)$.

It is interesting to note that there is essentially only one physical parameter $m$ that determines the properties of the classical spacetime, while the parameter $\gamma$ affects only the dynamical equations through Eqs.(\ref{eq3a}) - (\ref{eq3d}), but has no effect on the spacetime. This is true only classically, and quantum mechanically $\gamma$ does affect the properties of quantum spacetimes. In particular, consideration of black hole thermodynamics in LQG requires $\gamma \simeq 0.2375$ \cite{Meissner:2004ju}.

\subsection{
Effective Loop Quantum Black Holes
}

In LQC, leading-order quantum corrections are incorporated by introducing quantum parameters $\delta_b$ and $\delta_c$ which stand for \textit{edge length} of holonomy. This procedure is called polymerization \cite{Ashtekar:2005qt}. The resultant quantum-corrected spacetime can be obtained from the effective Hamiltonian. Because the Schwarzschild black hole interior can be treated as a KS cosmological model, polymerization can be used to construct an effective Hamiltonian that describes the interior of a quantum black hole.
Polymerization is realized by the replacements (\ref{eq0.1})
\bqn
\lb{eq8}
N^{\rm{eff}} &=& \frac{\gamma\delta_b \sqrt{p_c}}{\sin(\delta_b b)},\\
\lb{eq9}
H^{\rm{eff}}[N^{\rm{eff}}] &=& - \frac{1}{2\gamma G}\Bigg[2 \frac{\sin(\delta_c c)}{ \delta_c} p_c\nb\\
&& + \left(\frac{\sin(\delta_b b)}{ \delta_b} + \frac{\gamma^2\delta_b}{\sin(\delta_b b)}\right)p_b\Bigg].
\eqn
Note that in writing the above expressions, we assumed $p_c > 0$ without loss of generality, as shown above.
Eqs.\eqref{eq8} and \eqref{eq9} will reduce, respectively, to the classical expressions of Eqs.\eqref{eq1} and \eqref{eq2} in the limit $\delta_b \rightarrow 0$ and $\delta_c \rightarrow 0$.

Depending on the model, $\delta_b$ and $\delta_c$ can be chosen as constants or functions of the phase-space variables $(p_b, p_c, b, c)$.
Different choices correspond to different quantization schemes and lead to different effective dynamics. Some of them suffer from dependence on fiducial cell \cite{Ashtekar:2005qt,Chiou:2008nm}, and some suffer from large quantum effects in low curvature regions \cite{Boehmer:2007ket}. We list some of the models considered so far in Table \ref{scheme}, and for more details, see, for example, \cite{Olmedo:2016ddn,Ashtekar:2018cay,Ashtekar:2020ifw,Gambini:2022hxr,Ashtekar:2023cod} and references therein.

Assuming that  $\delta_b$ and $\delta_c$ depend only on $p_b$ and $p_c$ [cf. Table \ref{scheme}], the EoMs corresponding to the above effective Hamiltonian are given by
\bqn
\lb{eq2.25m}
\dot b&=& G \gamma  \frac{\partial H^{\rm{eff}}[N^{\rm{eff}}]}{\partial p_b}\nb\\
&=& -\frac{1}{2}\Bigg\{2 \left(\frac{c\cos(\delta_c c)}{\delta_c}-\frac{\sin(\delta_c c)}{\delta_c^2}\right) \frac{\partial \delta_c}{\partial p_b} p_c\\
&& +\left[\frac{\gamma^2 \delta_b}{\sin(\delta_b b)} +\frac{\sin(\delta_b b)}{\delta_b} \right]\nb\\
&& +p_b \frac{\partial}{\partial p_b} \left[\frac{\gamma^2 \delta_b}{\sin(\delta_b b)} +\frac{\sin(\delta_b b)}{\delta_b}\right]
\Bigg\},\\
\lb{eq2.25n}
\dot c&=&2G \gamma  \frac{\partial H^{\rm{eff}}[N^{\rm{eff}}]}{\partial p_c}\nb\\
&=& -\Bigg\{2 \left(\frac{c\cos(\delta_c c)}{\delta_c}-\frac{\sin(\delta_c c)}{\delta_c^2}\right) \frac{\partial \delta_c}{\partial p_c} p_c\nb\\
&& +2\frac{\sin(\delta_c c)}{\delta_c}\nb\\
&& +p_b \frac{\partial}{\partial p_c} \left[\frac{\gamma^2 \delta_b}{\sin(\delta_b b)} +\frac{\sin(\delta_b b)}{\delta_b} \right]
\Bigg\},\\
\lb{eq2.25o}
\dot p_c &=&-2G \gamma  \frac{\partial H^{\rm{eff}}[N^{\rm{eff}}]}{\partial c} \nb\\
&=& 2 p_c \cos(\delta_c c) , \\
\lb{eq2.25p}
\dot p_b &=&-G \gamma  \frac{\partial H^{\rm{eff}}[N^{\rm{eff}}]}{\partial b} \nb\\
&=& \frac{p_b}{2} \cos(\delta_b b) \left[1-\frac{\gamma^2 \delta_b^2}{\sin^2(\delta_b b)}\right].
\eqn
As mentioned above, when $|\delta_b|$ and $|\delta_c|$ are small,  $H^{\rm{eff}}[N^{\rm{eff}}]$ will reduce to $H^{\rm{GR}}[N^{\rm{GR}}]$. Conversely, the quantum effects become significant  when  $|\delta_b| \sim \mathcal{O}(1)$ and/or $|\delta_c| \sim \mathcal{O}(1)$ \cite{Boehmer:2007ket,Chiou:2008nm}.
In this paper, we mainly focus on the so-called $\bar\mu$ scheme first proposed in \cite{Boehmer:2007ket}. In LQC, as mentioned above, this is the only scheme that leads to a consistent quantum cosmological model \cite{Ashtekar:2006wn,Ashtekar:2011ni}.

\section{ B\"ohmer-Vandersloot Model}\lb{sec-bv}
\renewcommand{\theequation}{3.\arabic{equation}}\setcounter{equation}{0}

In LQC, a consistent prescription for the polymerization parameters was obtained by requiring the physical area $A_{x\theta} \left(= \delta_b\delta_c p_b\right)$ \cite{Ashtekar:2006wn} \footnote{In the homogeneous isotropic universe, we have $\delta_b  = \delta_c  \equiv \delta$. As a result, the condition (\ref{deltac}) uniquely determines $\delta$. However, in the KS spacetime, this is no longer the case. Therefore, in general, we have $\delta_b \not= \delta_c$, and now one more condition is needed to uniquely determine both $\delta_b$ and $ \delta_c$ uniquely.} of the closed holonomy loop in the $(x,\theta)$-plane be equal to the minimum area gap $\Delta$ predicted by LQG, so that
\bqn
\lb{deltac}
\delta_b\delta_c p_b = \Delta.
\eqn
However, for the holonomies on the 2-sphere, the loop does not close, and BV requires that
$A_{\theta\phi} \left(=\delta_b^2 p_c\right)$ be equal to the minimum area, i.e.
\bqn
\lb{deltab}
\delta_b^2 p_c = \Delta.
\eqn
Then, from the above equations, we obtain
\bq
\lb{eq3}
\delta_b = \sqrt{\frac{\Delta}{p_c}}, \quad
\delta_c = \frac{\sqrt{\Delta {p_c}}}{p_b},
\eq which are all dimensionless and are often referred to as  the $\bar\mu$-scheme for the spherically symmetric spacetimes \cite{Boehmer:2007ket}. Hence, we have
\bqn
\lb{eq2.25k}
\frac{\partial \delta_b}{\partial p_b}=0, \quad \frac{\partial \delta_b}{\partial p_c}=-\frac{\delta_b}{2p_c},\\
\lb{eq2.25l}
\frac{\partial \delta_c}{\partial p_b}=-\frac{\delta_c}{p_b},  \quad  \frac{\partial \delta_c}{\partial p_c}=\frac{\delta_c}{2p_c}.
\eqn
Inserting them into Eqs.(\ref{eq2.25m}) - (\ref{eq2.25p}), we obtain
\bqn
\lb{eq2.25s}
\dot b &=&-\frac{c \mathcal{F}}{2\frac{\sin(\delta_c c)}{\delta_c}}\cos(\delta_c c),\\
\lb{eq2.25t}
\dot c &=& -\frac{\sin(\delta_c c)}{\delta_c\mathcal{F}} \Bigg\{b\cos(\delta_b b)-b\cos(\delta_b b)\frac{\gamma^2 \delta_b^2}{\sin^2(\delta_b b)} \nb\\
&&  +2\frac{\gamma^2 \delta_b}{\sin(\delta_b b)} \nb\\
&& +\delta_c c\cot(\delta_c c) \left[\frac{\gamma^2 \delta_b}{\sin(\delta_b b)} +\frac{\sin(\delta_b b)}{\delta_b}\right] \Bigg\}, ~~~~~~~\\
\lb{eq2.25u}
\dot p_c &=& 2 p_c \cos(\delta_c c), \\
\lb{eq2.25v}
\dot p_b &=& \frac{p_b}{2} \cos(\delta_b b) \left[1-\frac{\gamma^2 \delta_b^2}{\sin^2(\delta_b b)}\right], ~~~~~~~~
\eqn where
\bq
\lb{eq2.25f}
\mathcal{F} \equiv \frac{\gamma^2 \delta_b}{\sin(\delta_b b)} +\frac{\sin(\delta_b b)}{\delta_b}.
\eq
After taking the following identity into account
\bq
\lb{eq2.25fA}
\frac{p_c}{p_b}=\frac{\delta_c}{\delta_b}=-\frac{\mathcal{F}}{2\frac{\sin(\delta_c c)}{\delta_c}},
\eq it can be shown that Eqs.\eqref{eq2.25s}-\eqref{eq2.25v} reduce to Eqs.(58)-(61) given in \cite{Boehmer:2007ket}. In particular, the effective Hamiltonian (\ref{eq9}) takes the form
\bqn
\lb{eq2.25fAa}
&& H^{\rm{eff}}[N^{\rm{eff}}] = - \frac{p_c}{2\gamma G \sin{(\delta_b b)} \delta_c} C_{\rm{BV}},\nb\\
\lb{eq2.25fAb}
&& C_{\rm{BV}} \equiv 2\sin{(\delta_b b)}\sin{(\delta_c c)}+ \sin{(\delta_b b)}^2+\gamma^2 \delta_b^2. ~~~~~~~~
\eqn

\subsection{
Quantum Effects Near the Classical Black Hole Horizon
}

The classical black hole horizon is  at
\bq
\lb{eq2.44}
T^{\rm{GR}}_H \equiv \ln(2m),
\eq as can be seen clearly from Eq.(\ref{eq6d}), at which we have $p_b^{\rm{GR}} = 0$.
Before solving the EoMs, let us first estimate the quantum effects near $T = T^{\rm{GR}}_H$.
Substituting the classical Schwarzschild black hole solution given by Eqs.\eqref{eq4a}-\eqref{eq4d} into Eq.\eqref{eq3}, we find
\bqn
|\delta_b|= \frac{\sqrt{\Delta}}{e^T}\rightarrow \frac{\sqrt{\Delta}}{2m}, \quad   T \rightarrow T^{\rm{GR}}_H, ~~~~
\eqn but
\bqn\lb{bv-delta_c}
|\delta_c|= \frac{\sqrt{\Delta}}{\sqrt{2m e^{-T}-1}} \rightarrow \infty, \quad   T \rightarrow T^{\rm{GR}}_H.
\eqn
Eq.\eqref{bv-delta_c} indicates that the BV solution has large quantum effects near $T \simeq T^{\rm{GR}}_H$ \cite{Boehmer:2007ket}. To characterize these effects, in the vicinity of $T^{\rm{GR}}_H$ let us introduce $\epsilon$ through $T=T_{\epsilon}\equiv T^{\rm{GR}}_H(1-\epsilon)$ with $\epsilon \ll 1$. Then, assuming that  at $T_{\epsilon}$ we have $|\delta_c\left(T_{\epsilon}\right)| \simeq \mathcal{O}(1)$, so that
\bqn
\left|\delta_c\right|_{T=T_{\epsilon}}&=& \frac{\sqrt{\Delta}}{\sqrt{2m e^{-T_{\epsilon}}-1}}
\nb\\
&\approx &  \frac{\sqrt{\Delta}}{\sqrt{(2m)^{\epsilon}-1}} \sim \mathcal{O}(1),
\eqn which leads to
\bqn\lb{epsilon}
\epsilon \simeq \frac{\ln(\Delta +1)}{\ln(2m)}.
\eqn
From this expression, we can see that $\epsilon$ is small and that as $m$ increases, $\epsilon$ decreases. That is to say, we need to get very close to $T^{\rm{GR}}_H$ in order to see the quantum effects for massive black holes. In fact, in the following, we will show that such quantum effects are so large that black/white hole horizons are never formed. Recall that the geometric radius $r$ of the two spheres $T, \; x =$ constant now is given by $r = e^T$. This in turn implies that in the BV approach, quantum black hole horizons do not exist in the entire region $T > T_{\cal{T}}$.

\subsection{
Initial Conditions
}

To demonstrate our above claim, let us first consider the initial conditions.  {
Since Eqs.(\ref{eq2.25s}) - (\ref{eq2.25v}) are four first-order ordinary differential equations, four initial conditions are generally needed. However, these initial conditions must also satisfy the
Hamiltonian constraint
\bq
\lb{C_BV}
H^{\rm{eff}} (T) = 0,
\eq which also holds at $T = T_i$. Therefore, only three are independent. As a result, the phase space of the initial conditions is three-dimensional (3D). Without loss of generality, we can first choose the initial conditions for $p_b(T_i)$,  $p_c(T_i)$ and $b(T_i)$ at the initial time $T = T_i$, and then solve the effective Hamiltonian constraint to find $c(T_i)$. It is clear that the obtained 3D phase space includes all the possible real values of $p_b(T_i)$,  $p_c(T_i)$ and $b(T_i)$. However, to compare the resultant BV spacetimes with the Schwarzschild spacetimes, we choose them as their corresponding values of GR, that is,
\bqn
\lb{GRVsA}
&& p_b(T_i) = p^{\rm{GR}}_b(T_i),  \quad p_c(T_i) = p^{\rm{GR}}_c(T_i), \nb\\
&& b(T_i) = b^{\rm{GR}}(T_i),
\eqn and
\bqn
\lb{GRVsB}
H^{\rm{eff}}(T_i) =0 \quad \Rightarrow \quad
c(T_i) = c^{\rm{eff}}(T_i).
\eqn
Once the initial conditions are chosen at a given initial time $T_i$, the EoMs Eqs.\eqref{eq2.25s}-\eqref{eq2.25v} will uniquely determine the four physical variables $\left(p_b, b; p_c, c\right)$ at any given later time $T$. }

{Due to the large quantum effects near $T^{\rm{GR}}_H$ as estimated in the last subsection, normally we choose $T_i$  far from $T^{\rm{GR}}_H$, that is, $T_i \ll T^{\rm{GR}}_H$ \footnote{
{It should be noted that $c(T_i) = c^{\rm{eff}}(T_i)$ obtained from $H^{\rm{eff}}(T_i) =0$ can be significantly different from $c^{\rm{GR}}(T_i)$, when $T_i \simeq T^{\rm{GR}}_H$, even $p_b(T_i)$,  $p_c(T_i)$ and $b(T_i)$ are chosen as their corresponding values of GR, as given by Eq.(\ref{GRVsA}), because now $H^{\rm{eff}}(T_i) \not= H^{\rm{GR}}(T_i)$. In fact, from Eq.\eqref{bv-delta_c} we find that $|\delta_c c|$ is very large near $T^{\rm{GR}}_H$, precisely because the fact that now $c(T_i)$ deviates from its classical value significantly. Then, the solution cannot be approximated by the classical Schwarzschild black hole at $T^{\rm{GR}}_H$, as shown in Table \ref{ts2} for $T_i = 0.685$. Recall that the Schwarzschild black hole horizon is located at $T_{H}^{\rm{GR}} \simeq 0.693$ for $m = \ell_{pl}$.}}.
On the other hand, near the throat $T = T_{\cal{T}}$ it is expected that the spacetime geometry will be dramatically different from that of GR, so the initial conditions chosen as those given by Eq.(\ref{GRVsA}) near $T_{\cal{T}}$ might lead the equation $H^{\rm{eff}}(T_i) = 0$ to a complex value of $c(T_i)$, as shown in Table \ref{ts2} for $T_i = -1.45$. Recall that the transition surface is located at $T_{\cal{T}} \simeq -1.49$ for $m = \ell_{pl}$. Therefore, in general, we choose $T_i$ so that
\bqn
\lb{Ti}
T_{\cal{T}} \ll T_i \ll T^{\rm{GR}}_H.
\eqn
To further understand the above arguments, we consider the initial conditions at different times $T_i
$'s. We compare the effective value of $c(T_i) = c^{\rm{eff}}(T_i)$ obtained from Eq.\eqref{GRVsB} with its corresponding classical value $c^{\rm{GR}}(T_i)$ at different times $T_i$'s in Table  \ref{ts2} for $m = \ell_{pl}$, for which we have $T_{\cal{T}} \simeq -1.49,$ and $T^{\rm{GR}}_H \simeq 0.693$.
From the Table we can see that Eq.\eqref{GRVsB} has no real-value solutions for $c(T_i)$ when $T_i$ is very close to either the transition surface $T_{\cal{T}}$ or to the classical horizon $T^{\rm{GR}}_H$, which means that the quantum effects are so large near these points that the effective Hamiltonian constraint Eq.\eqref{GRVsB}
has no physical solutions for the chosen $p_b(T_i)$,  $p_c(T_i)$ and $b(T_i)$.
On the other hand, when far away from these points, the difference between  $c(T_i)$ and $c^{\rm{GR}}(T_i)$ is small. Therefore, in the following, we choose $T_i$ so that the condition (\ref{Ti}) is always satisfied.}

\subsection{
Numerical Results
}

Once the initial time and conditions are specified, we are ready to solve the EoMs
(\ref{eq2.25s}) - (\ref{eq2.25v}) numerically. To monitor the numerical errors, we should closely pay attention to the effective Hamiltonian  given by Eq.(\ref{eq2.25fAb}), which is required to vanish identically along any of the physical trajectories. However, numerically, this is true only up to a certain accuracy. To ensure that such numerical calculations are reliable and that our physical conclusions will not depend on these numerical errors, we run our {\em Mathematica} code in supercomputers with high precision. In particular, in all calculations, we require that the Working Precision and Precision Goal be respectively 250 and 245, respectively, where Working Precision specifies how many digits of precision should be maintained in internal computations of {\em Mathematica}, and Precision Goal specifies how many effective digits of precision should be sought in the final result.

\subsubsection{Classical Results}

Before solving the dynamical equations of the BV model,  Eqs.\eqref{eq2.25s}-\eqref{eq2.25v}, let us first consider their classical limits, which can be obtained by taking $\delta_{b, c} \rightarrow 0$. Then, the reduced equations have the relativistic exact solution given by Eq.(\ref{eq6d}). Setting   $T = T_i = 0.3$, Eq.(\ref{eq6d}) will give us the corresponding values, $b^{\rm{GR}}(T_i), p_b^{\rm{GR}}(T_i), c^{\rm{GR}}(T_i)$ and $p_c^{\rm{GR}}(T_i)$. Taking these values as the initial conditions, using our numerical code, we solve the corresponding classical equations of Eqs.\eqref{eq2.25s}-\eqref{eq2.25v}, obtained by setting $\delta_b = \delta_c = 0$. In Fig. \ref{m1p250T03classical} we plot all the dynamic variables $b^{(n)}, p^{(n)}_b, c^{(n)}, p^{(n)}_c$ together with their exact values of Eq.(\ref{eq6d}), denoted by $b^{(a)}, p^{(a)}_b, c^{(a)}, p^{(a)}_c$. From the figure, it can be seen that our numerical results are indistinguishable from the exact ones within our numerical errors. In this figure, we also plot the quantities $N_{\lambda}N^{\lambda}$ and $\Theta_{\pm}$, from which it can be seen clearly that both of them go to zero as $T \rightarrow T^{\rm{GR}}_{H} = \ln(2m)$, where $N_{\lambda}$ is defined by Eq.(\ref{nvector}). Note that in these plots, we require the Working Precision and Precision Goal to be 100, which is already sufficiently accurate for the current case. With the confidence of our numerical code, let us turn to solve Eqs.\eqref{eq2.25s}-\eqref{eq2.25v} for the effective BV Hamiltonian.

\begin{figure*}[htbp]
$\begin{array}{cc}
\includegraphics[width=0.45\textwidth]{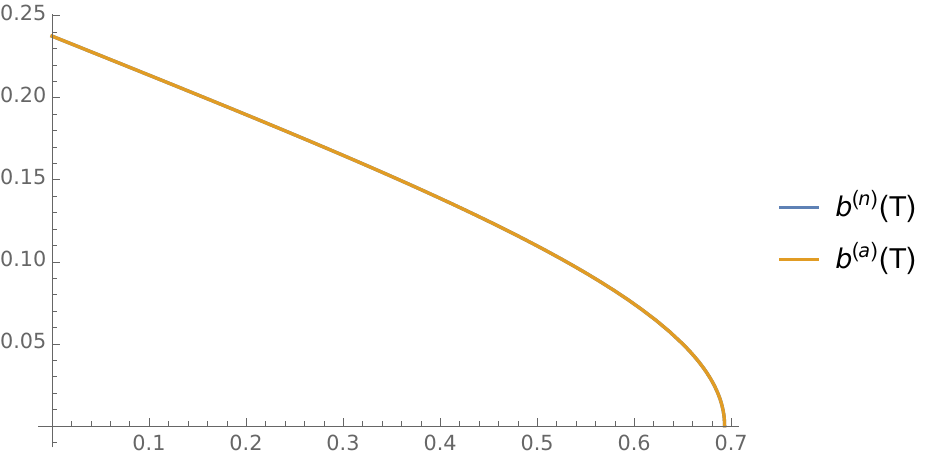}&
\includegraphics[width=0.45\textwidth]{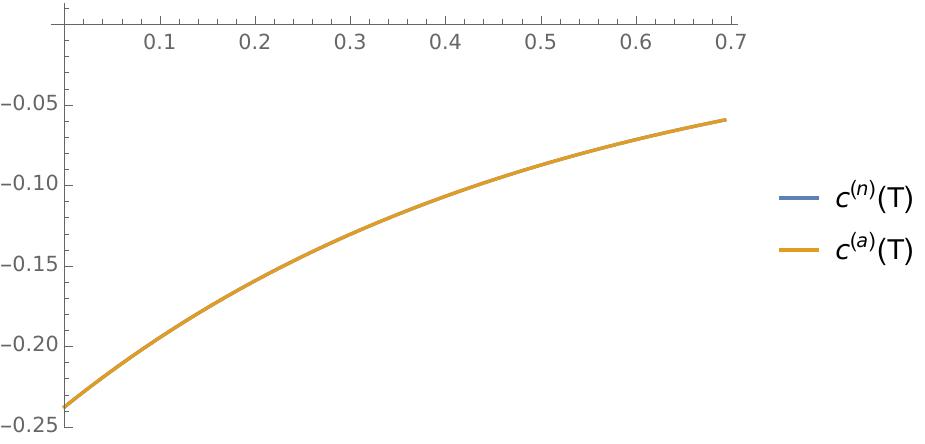}\\
\\[6pt]
\\
\includegraphics[width=0.45\textwidth]{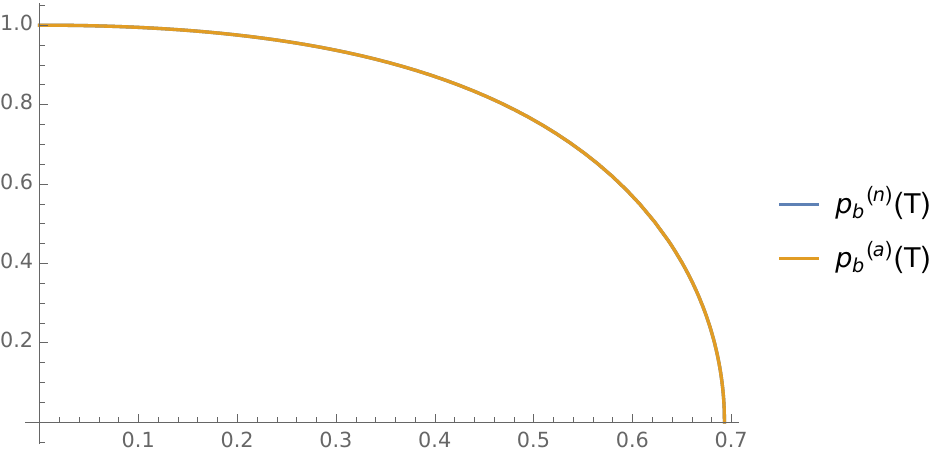}&
\includegraphics[width=0.45\textwidth]{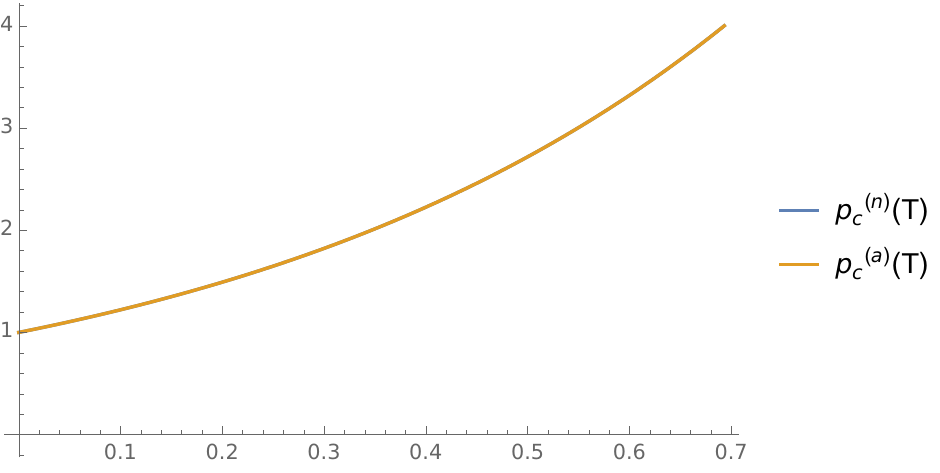}\\
\\[6pt]
\\
	\includegraphics[width=0.45\textwidth]{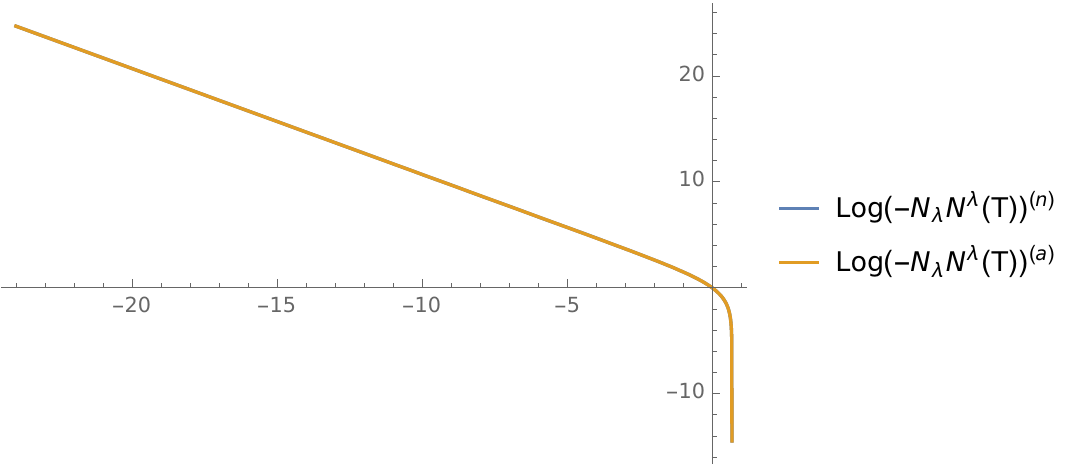}&
\includegraphics[width=0.45\textwidth]{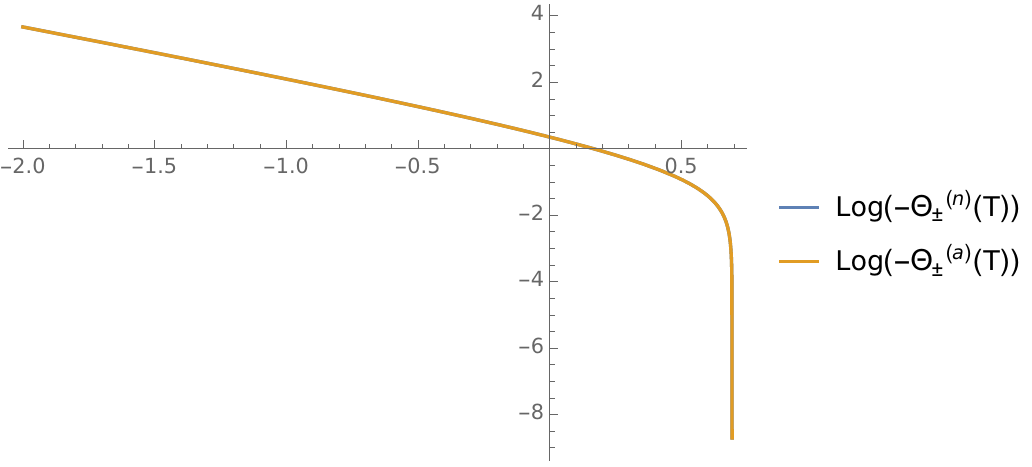}\\
\\[6pt]
\\
\includegraphics[width=0.45\textwidth]{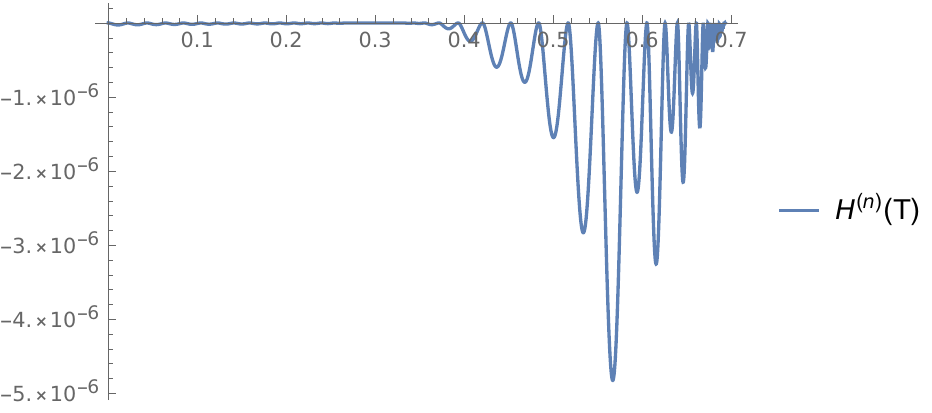}&\\
\\[6pt]
\end{array}$
\caption{Plots of  the four classical physical variables $\left(b, c, p_b, p_c\right)$ and the classical  Hamiltonian  defined by Eq.(\ref{eq2}) vs $T$ (denoted by the horizontal line in each sub-figure). The superscript $(n)$ represents a numerical value, while the superscript $(a)$ represents an analytical value. Here, we require that the Working Precision and Precision Goal be 100. From these figures it can be seen that the numerical solutions are indistinguishable from the exact ones within our numerical errors. The mass parameter $m$ is chosen as  $m/\ell_{pl}=1$, for which we have  $T_{\cal{T}} \simeq -1.49$ and $T^{\rm{GR}}_H \approx0.693$. The initial time is set at $T_i = 0.3$.}
\lb{m1p250T03classical}
\end{figure*}

\subsubsection{Asymptotic Behavior of Spacetimes as $T \gg T_{\cal{T}}$}

With the above in mind, let us first consider the case $m = \ell_{pl}$.
In this case, we have $T_{\cal{T}} \simeq -1.49$ and $T^{\rm{GR}}_H \simeq 0.693$, as noted above.
The initial moment is chosen as $T_i = 0.3$, which satisfies Eq.(\ref{Ti}). From Table \ref{ts2} we can also see that at this point $c(T_i)$ is very close to its corresponding classical value $c^{\rm{GR}}(T_i)$. In Fig. \ref{m1p250T03} we plot all four variables $b, \; c,\; p_b$ and $p_c$, together with $H^{\rm{eff}}$, the corresponding Kretchmann scalar $K$ and the metric components $N^2$ and $g_{xx}$, where the Kretchmann scalar $K$ is defined as $K(T) \equiv R_{\alpha\beta\mu\nu} R^{\alpha\beta\mu\nu}$.
From Fig. \ref{m1p250T03} (h), we can see that the maximal errors occur near
$T \simeq 10$ at which $\left|H^{\rm{eff}}\right| \le 2.0 \times 10^{-8}$. Before or after it, we have $\left|H^{\rm{eff}}\right| \ll 10^{-8}$. Therefore, our numerical computations are reliable.

\begin{figure*}[htbp]
\begin{tabular}{cc}
\includegraphics[width=0.45\textwidth]{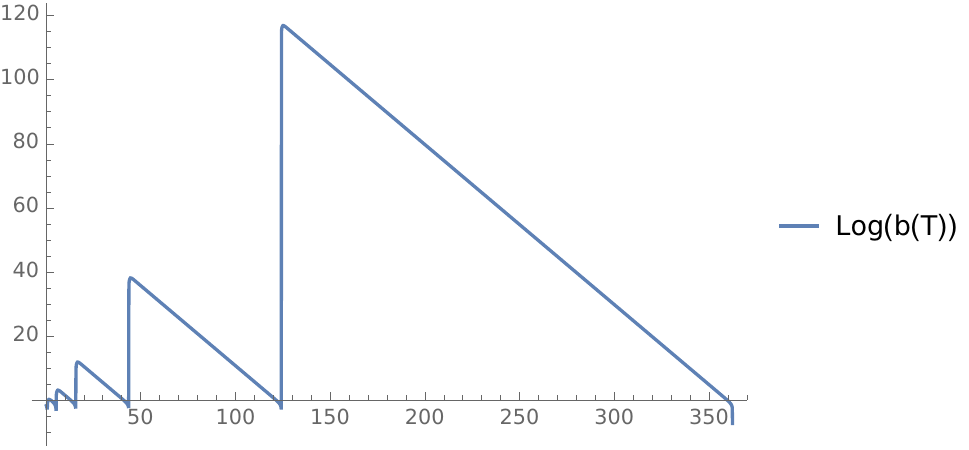}&
\includegraphics[width=0.45\textwidth]{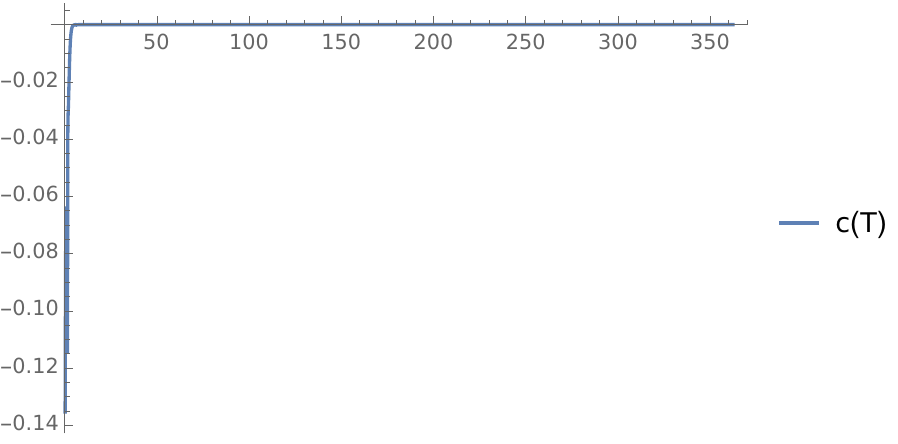}\\
(a) & (b) \\[6pt]
\\
\includegraphics[width=0.45\textwidth]{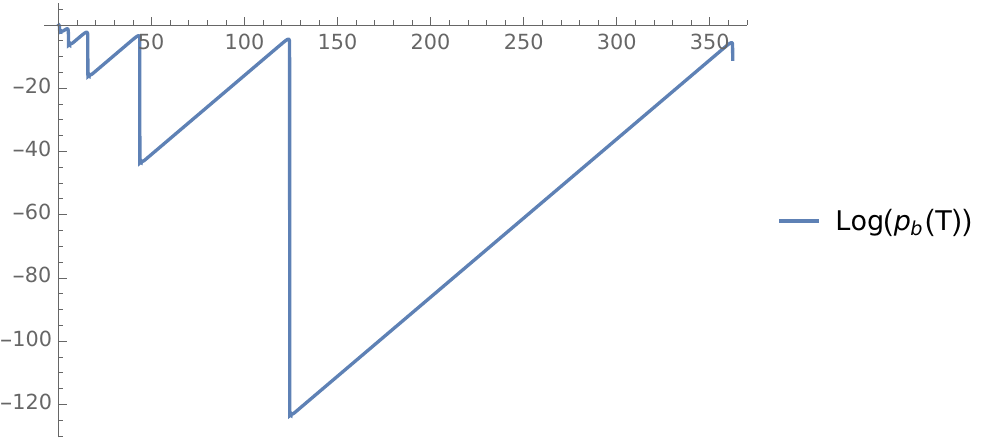}&
\includegraphics[width=0.45\textwidth]{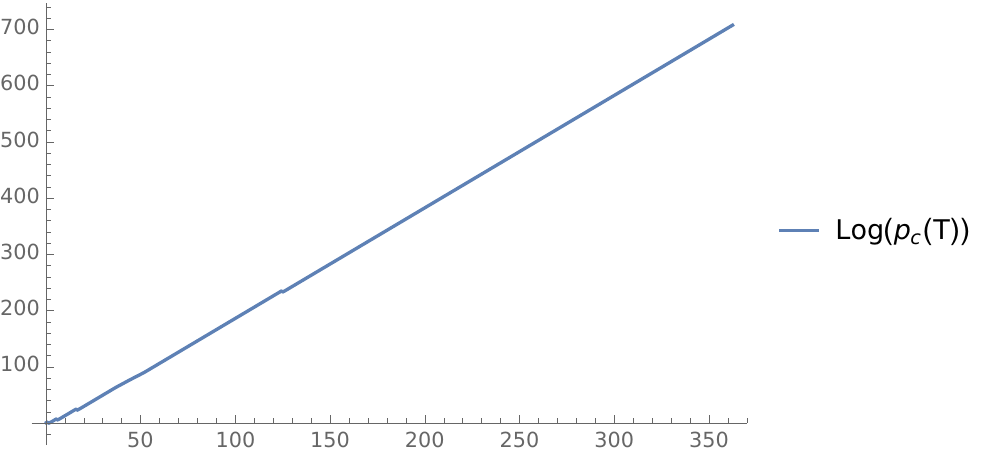}\\
(c) & (d) \\[6pt]
\\
	\includegraphics[width=0.45\textwidth]{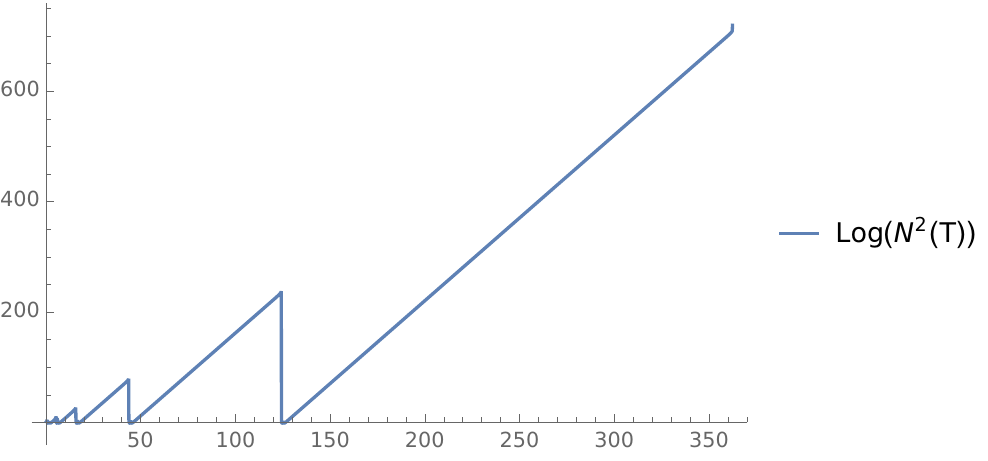}&
\includegraphics[width=0.45\textwidth]{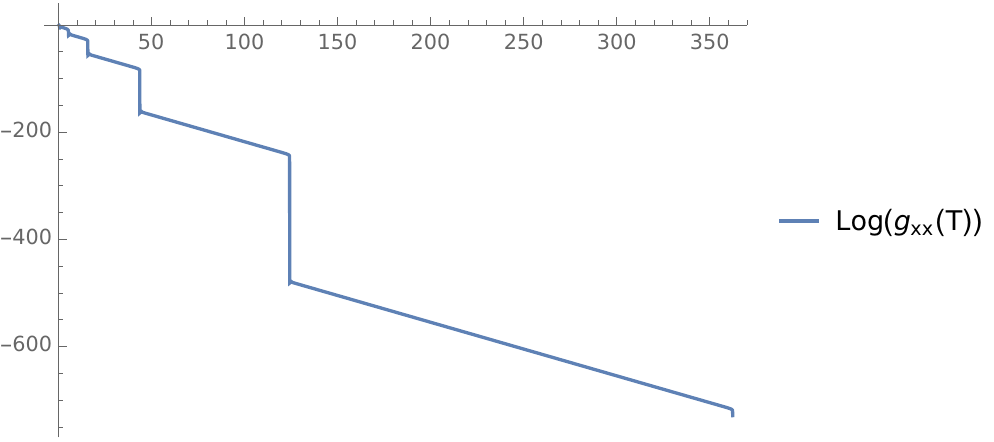}\\
(e) & (f)   \\[6pt]
\\
\includegraphics[width=0.45\textwidth]{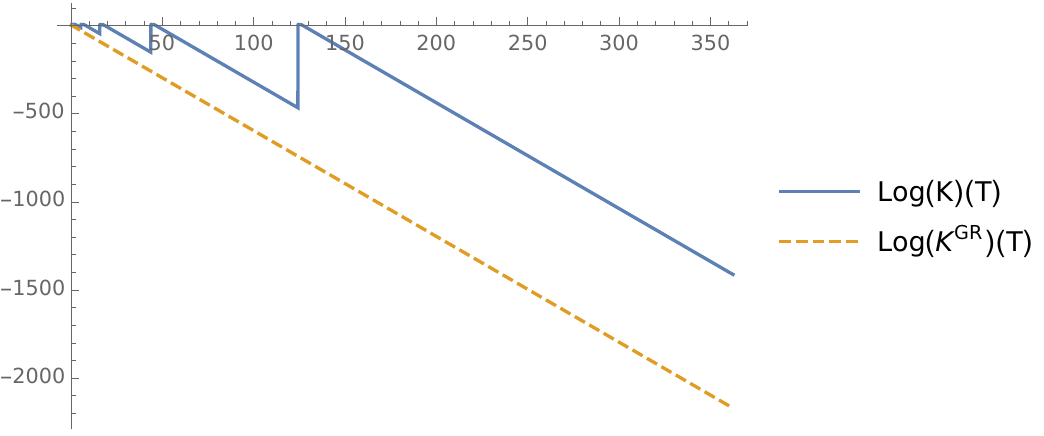}&
\includegraphics[width=0.45\textwidth]{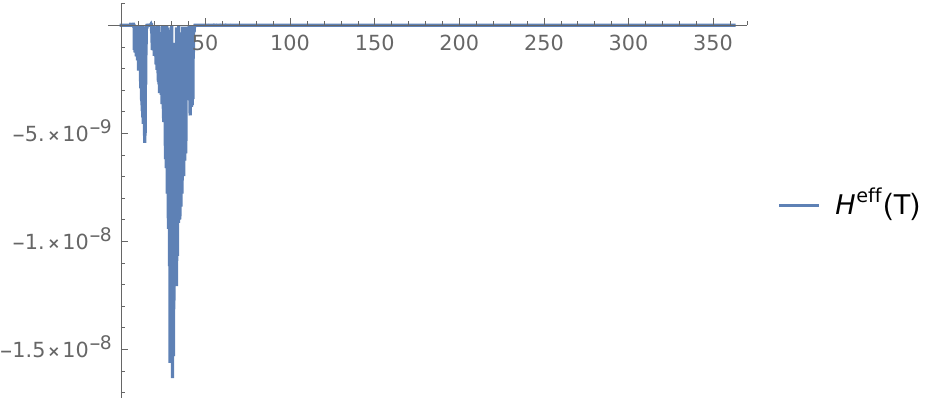}\\
	(g) & (h)  \\[6pt]
	\end{tabular}
\caption{Plots of  the four physical variables $\left(b(T), c(T), p_b(T), p_c(T)\right)$ given respectively by the sub-figures
(a), (b), (c) and (d), and the metric components $N^2(T)$, $g_{xx}(T)$, the Kretschmann scalar $K(T)$, and the effective Hamiltonian  $H^{\text{eff}}(T)$ defined by Eq.(\ref{eq2.25fAb}), given respectively by the sub-figures (e), (f), (g) and (h). In the sub-figure (g), we also plot 
the classical counterpart $K^{\rm{GR}}(T) \equiv 48m^2/p_c^3$ of the Kretschmann scalar. These quantities are all functions of $T$ only, and in each of the sub-figures, the $T$-axis is the horizontal line. The mass parameter $m$ is chosen as  $m/\ell_{pl}=1$, for which we have  $T_{\cal{T}} \simeq -1.49$ and $T^{\rm{GR}}_H \approx0.693$. The initial time is set at $T_i = 0.3$.}
\lb{m1p250T03}
\end{figure*}

\begin{figure*}[htbp]
\begin{tabular}{cc}
\includegraphics[width=0.45\textwidth]{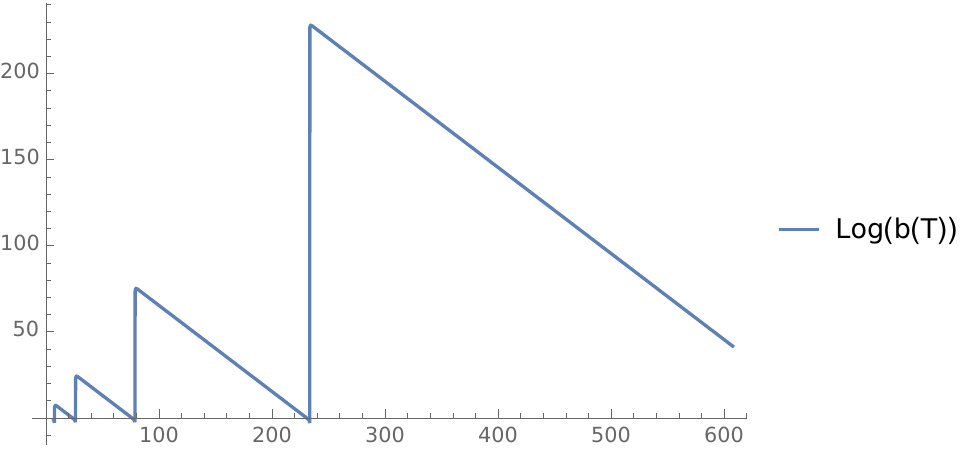}&
\includegraphics[width=0.45\textwidth]{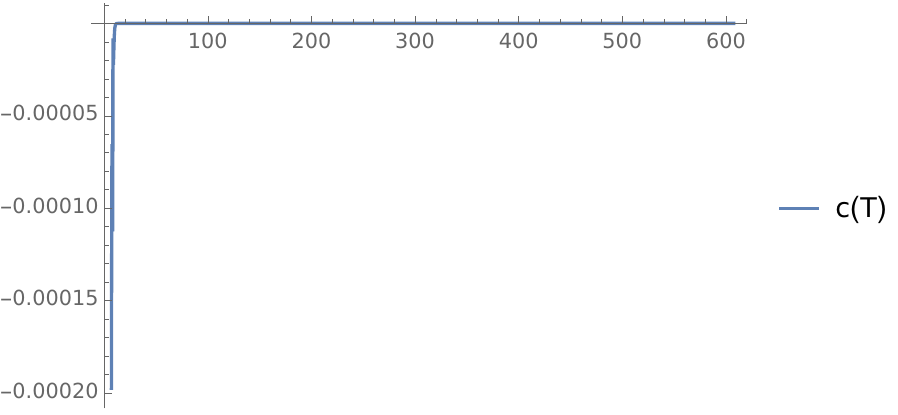}\\
\\[6pt]
\\
\includegraphics[width=0.45\textwidth]{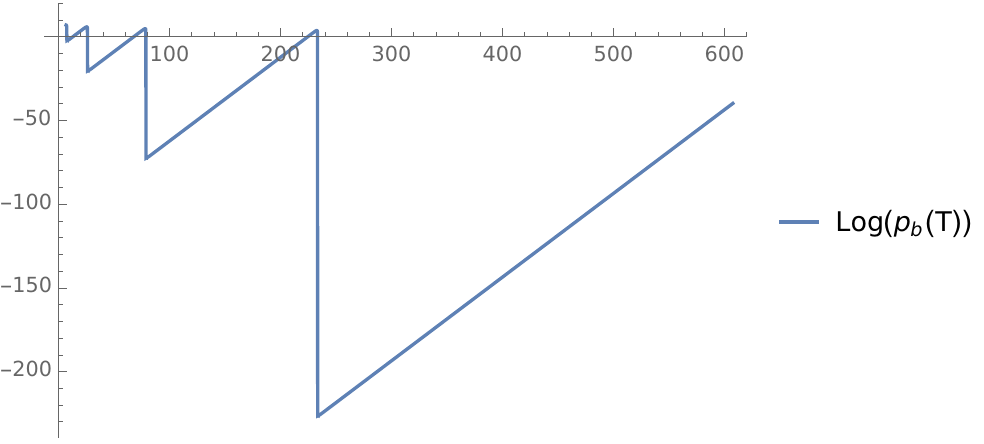}&
\includegraphics[width=0.45\textwidth]{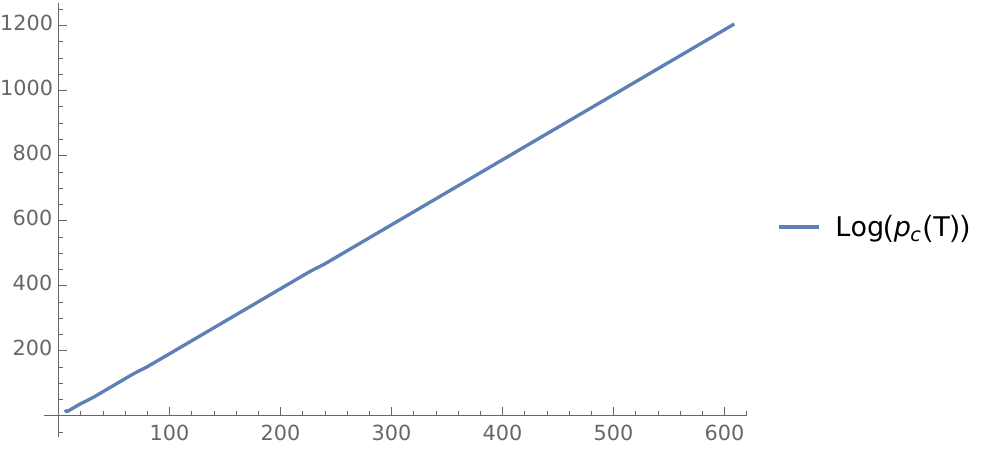}\\
\\[6pt]
\\
	\includegraphics[width=0.45\textwidth]{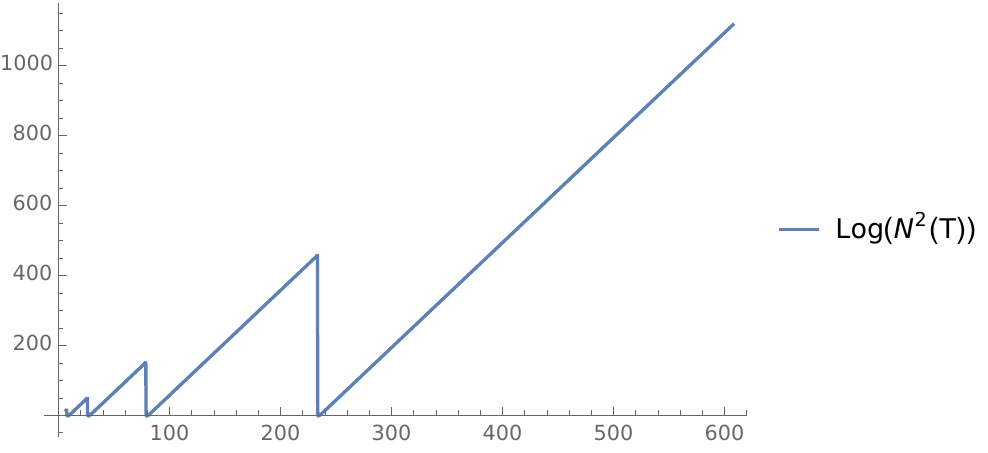}&
\includegraphics[width=0.45\textwidth]{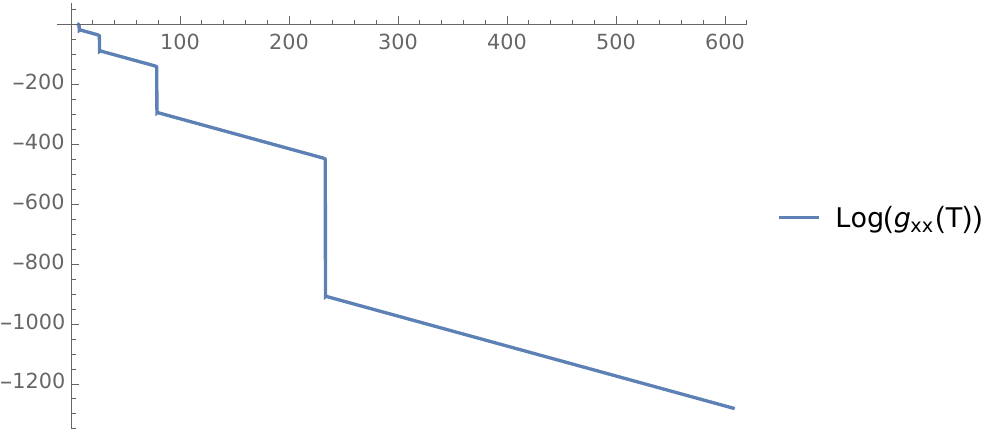}\\
\\[6pt]
\\
\includegraphics[width=0.45\textwidth]{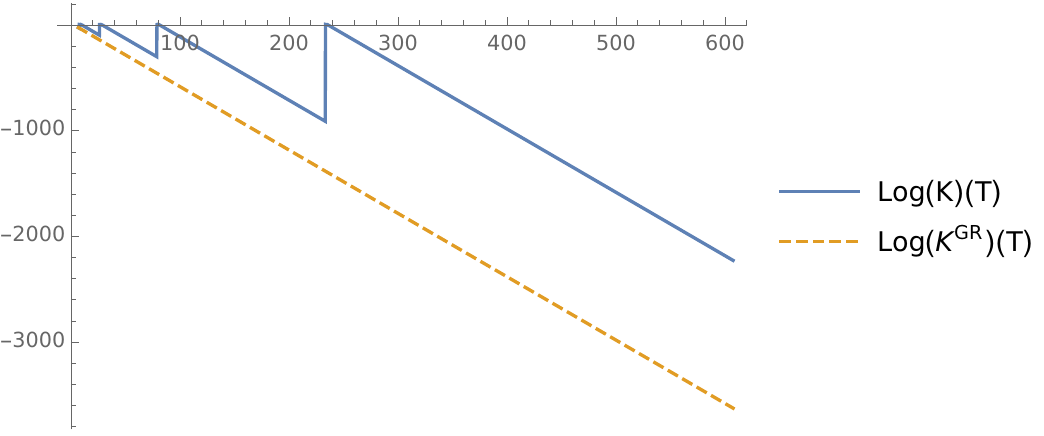}&
\includegraphics[width=0.45\textwidth]{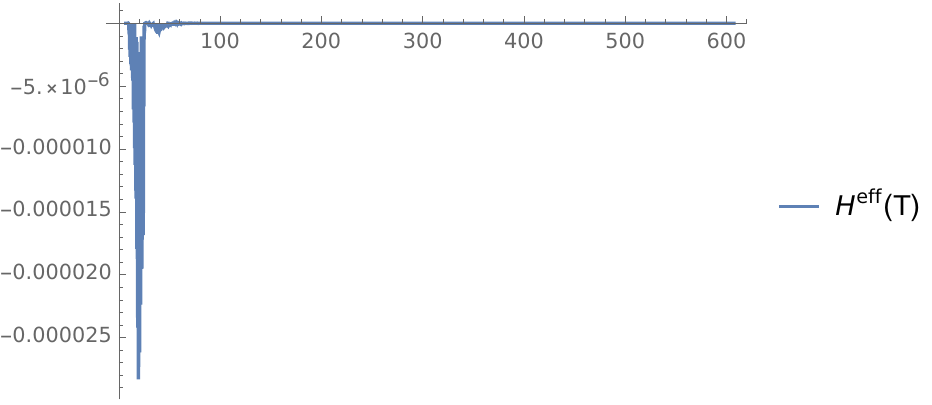}\\
\\[6pt]
	\end{tabular}
\caption{Plots of  the four physical variables $\left(b, c, p_b, p_c\right)$ and  $H^{\rm{eff}}$, together with the metric components $N^2$, $g_{xx}$,  the Kretchmann scalar $K$, and the classical counterpart $K^{\rm{GR}}(T) \equiv 48m^2/p_c^3$   with  $m/\ell_{pl}=10^3$, for which we have  $T_{\cal{T}} \simeq 0.8327$  and $T^{\rm{GR}}_H \approx 7.6009$. The initial time is set at $T_i = 7.0$.   In each of the subfigures, the horizontal line represents the $T$-axis.}
\lb{m10to3p250T7}
\end{figure*}

On the other hand, from Fig. \ref{m1p250T03} (d), we can see that the geometric radius $r = \sqrt{p_c}$ is exponentially increasing. As a result, the metric coefficient $g_{xx}$ decreases exponentially, but never be zero precisely at any given (finite) time $T$, as shown in Fig. \ref{m1p250T03} (f).
In addition, the square of the lapse function $N^2$ is also oscillating with a similar period as that of $p_b$, but after each circle of oscillations, it becomes increasingly larger [cf. Fig. \ref{m1p250T03} (e)]. However, it always remains finite. Moreover, in Fig. \ref{m1p250T03} (g), we plot the
Kretchmann scalar $K(T)$ together with its classical counterpart $K^{\rm{GR}}(T) \equiv 48m^2/p_c^3$, from which we can clearly see that the quantum geometric effects indeed become very large near the classical black hole horizon $T \simeq T^{\rm{GR}}_H$. This deviation lasts for quite a while,
$T \in \left(T^{\rm{GR}}_H, 160\right)$ for $m = \ell_{pl}$.

The above results are sharply in contrast to the classical case, in which $\left. N^2\right|_{\rm{GR}} = e^{2T}/(2me^{-T} -1) \rightarrow \infty$ at the black hole horizon, $T^{\rm{GR}}_H = \ln  2 \simeq 0.693$, while $p_b^{\rm{GR}}$ becomes zero  precisely at $T^{\rm{GR}}_H$, so is $g_{xx}^{\rm{GR}}$, as can be seen from Eqs.(\ref{eq4d}) and (\ref{eq5}).

It should also be noted that similar results can be obtained when the initial time is chosen to be at $T_i = \ln(2m) - 0.05 \simeq 0.643$, which is also a point at which the difference between $c(T_i)$ and  $c^{\rm{GR}}(T_i)$ is negligible, as shown in Table \ref{ts2}. In addition, we consider the case $m = 10^{3} \ell_{pl}$. The corresponding physical quantities are plotted in Fig.   \ref{m10to3p250T7}.
From these figures, we can see that the metric coefficients, $\left(N^2, \; g_{xx}, \; p_c\right)$,  are all finite and non-zero for any given finite time $T > T_{\cal{T}}$.

{The above results can be understood as follows: From the condition (\ref{deltac}), we find 
\bq
\lb{eq.3.24}
\delta_c p_b = \sqrt{\Delta p_c}.
\eq
Thus, for any given finite time $T$, the right-hand side is always finite and non-zero. Then, if $p_b = 0$ at a time, say, $T_H$, we must have
$\delta_c(T_H) = \infty$, which implies that the quantum geometric effects become infinitely large. As a result,   $p_b$ will never be zero within a finite time.}

{To study further the asymptotic behaviors of the spacetimes, let us first notice that  $\log \left(N^2\right)$ and   $\log\left(g_{xx}\right)$ change periodically, but during each period $\log \left(N^2\right)$
increases almost linearly, while  $\log\left(g_{xx}\right)$
Decreases almost linearly. In contrast, $\log \left(p_c\right)$
Increases almost linearly all the time [See the analysis to be given below]. Therefore, during each period, we can approximate each of them as $F = F_0 T^{\alpha}$, where we find that
\bqn
\lb{eq.3.24a}
N^2 & \simeq& A_0 e^{3T}, \quad
g_{xx}   \simeq  B_0 e^{-T},  \quad p_c = p_c^{(0)} e^{2 T}, ~~~~
\eqn where $A_0$, $B_0$ and $p_c^{(0)}$ are constants, usually depending on the period considered, but the slopes remain almost constant. Then, we find that the effective energy-momentum tensor calculated from $\kappa^2 T_{\mu\nu} \equiv G_{\mu\nu}$ can be written as follows:
\bqn
\lb{eq3.24b}
\kappa^2 T_{\mu\nu} = \rho u_{\mu}u_{\nu} + p_x x_{\mu}x_{\nu} +
p_{\bot}\left(\theta_{\mu} \theta_{\nu} +\phi_{\mu} \phi_{\nu}\right), ~~~~~
\eqn with
\bqn
\lb{eq3.24c}
\rho  \simeq  p_x \simeq \frac{1}{p_c} + {\cal{O}}\left(e^{-3T}\right), \;\;\;
p_{\bot} \simeq {\cal{O}}\left(e^{-3T}\right),
\eqn where ($u_{\mu}, \; x_{\mu}, \; \theta_{\mu}, \; \phi_{\mu}$) are the unit vectors along the ($dT, dx, d\theta, d\phi$)-directions. In addition, we also have
\bqn
\lb{eq3.24d}
R   &\simeq& \frac{2}{p_c}, \quad
K  \simeq
\frac{4}{p_c^2}, \quad
C^{\alpha\beta\mu\nu}  C_{\alpha\beta\mu\nu}  \simeq
\frac{4}{3p_c^2}, ~~~
\eqn where $C_{\alpha\beta\mu\nu}$ denotes the Weyl tensor.
Thus, spacetime becomes asymptotically flat, but with the
Kretchmann scalar decreasing  as $p_c^{-2}$, quite similar to the case studied in \cite{Ashtekar:2018cay,Ongole:2022rqi,Ongole:2023pbs},
Instead of $p_c^{-3}$ as in the classic case. }

\begin{figure*}[htbp]
\begin{tabular}{cc}
\includegraphics[width=0.45\textwidth]{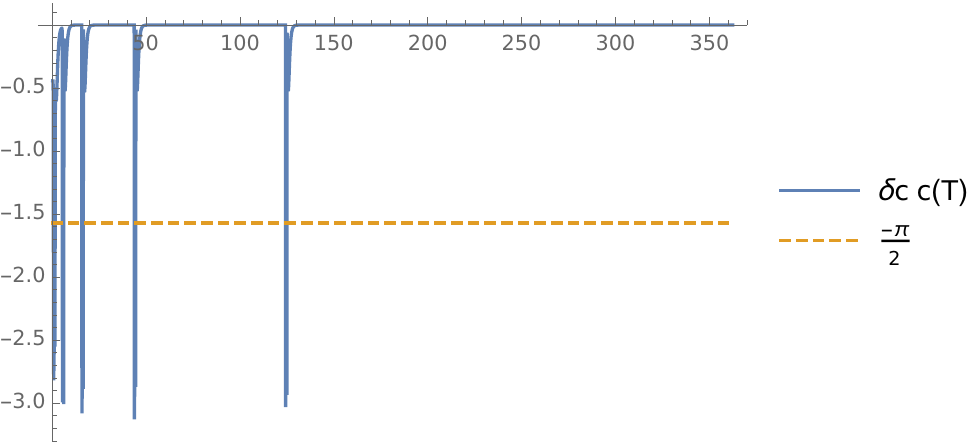}&
\includegraphics[width=0.45\textwidth]{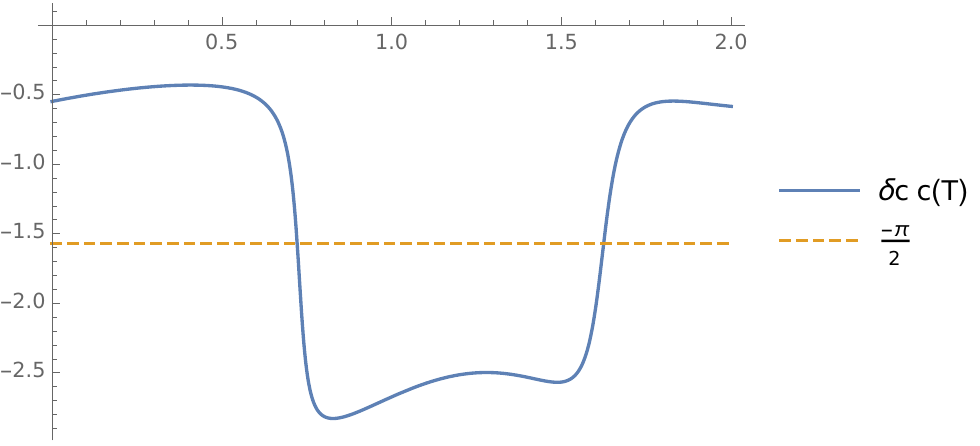}\\
(a) & (b) \\[6pt]
\\
\includegraphics[width=0.45\textwidth]{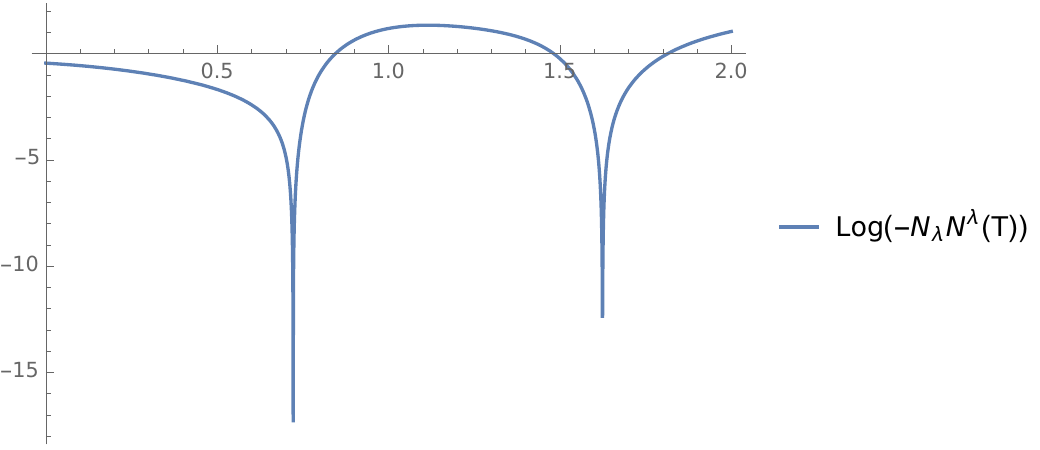}&
\includegraphics[width=0.45\textwidth]{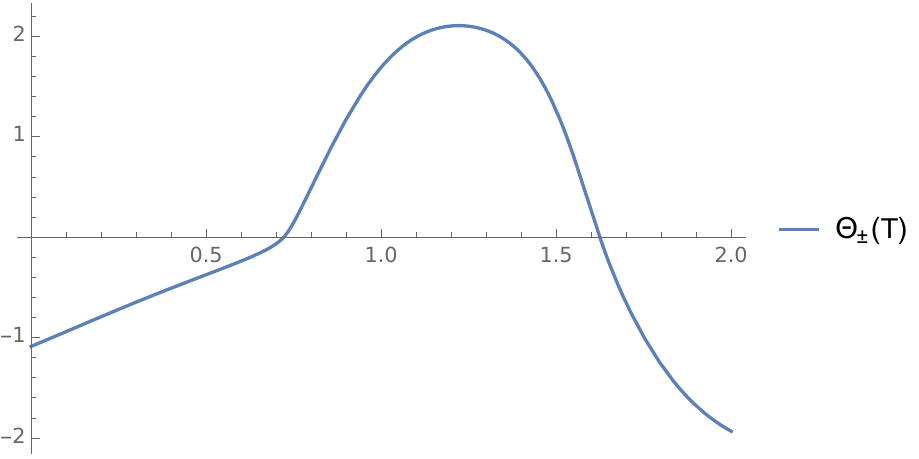}\\
(c) & (d)
\end{tabular}
\caption{Plots of  $\delta_c c$ and $\ln(-N_{\lambda}N^{\lambda})$ defined by Eq.(\ref{nvectorb}) in two different ranges of the coordinate $T$, represented by the horizontal lines. In particular,
the sub-figures Fig. \ref{fig6} (a) and Fig. \ref{fig6} (b) are the plots of $\delta_c c$ and the ones Fig. \ref{fig6} (c) and Fig. \ref{fig6} (d) are the plots of $\ln(-N_{\lambda}N^{\lambda})$. 
The mass parameter $m$ is chosen as  $m/\ell_{pl}=1$, for which we have  $T_{\cal{T}} \simeq -1.49$ and $T^{\rm{GR}}_H \approx0.693$. The initial time is chosen at $T_i = 0.3$.  Fig. \ref{fig6} (c) clearly shows that there exist two points at which $N^{\mu}$ becomes a null vector, $N_{\lambda}N^{\lambda} = 0$ in the interval $T \in (0, 2)$. On the other hand, Fig. \ref{fig6} (d) shows that the expansions change the sign across the transition surface.}
\lb{fig6}
\end{figure*}

\subsubsection{Non-existence of Black/White Hole Like Horizons}

{To see if a black or white hole like horizon exists, let us first consider if {\em a marginally trapped surface} will be developed in the BV model as $T$ (or $p_c$) increases. To achieve this goal, we can calculate the expansions of the in-going and out-going radially moving light rays
\cite{Hawking:1973uf,Ashtekar:2018cay,Wang:2003bt,Wang:2003xa,Hayward:1999ek}.  In particular, by introducing the unit vectors, $u_{\mu} \equiv N \delta^T_{\mu}$ and $s_{\mu} \equiv \sqrt{g_{xx}}\delta^x_{\mu}$, we construct two null vectors $\ell_{\mu}^{\pm} = \left(u_{\mu} \pm s_{\mu} \right)/\sqrt{2}$, which define the in-going and out-going radially moving light rays. The expansions of these light rays are given by \cite{Ashtekar:2018cay}
\bq
\lb{Eps}
\Theta_{\pm} \equiv m^{\mu\nu}\nabla_{\mu}\ell_{\nu}^{\pm} = - \frac{p_{c, T}}{\sqrt{2} N p_c},
\eq where $m_{\mu\nu} \equiv g_{\mu\nu} + u_{\mu}u_{\nu} - s_{\mu}s_{\nu}$. }

Note that the existence of a marginally trapped surface, at which we have $\Theta^{+} \Theta^{-} = 0$ \cite{Hawking:1973uf,Wang:2003bt,Wang:2003xa}, can be equally characterized by the vanishing of the norm of the normal vector to the 2-spheres, $T, \; x = $ Constant  \cite{Gong:2007md,Hayward:1999ek}.  In fact, introducing the normal vector to the surface
$\sqrt{p_c} = r_0$
\bqn
\lb{nvector}
N_{\mu} \equiv \frac{\partial (\sqrt{p_c}  - r_0)}{\partial x^{\mu}} = \frac{p_{c,T}}{2\sqrt{p_c}}\delta^T_{\mu},
\eqn where $r_0$ is a constant, we find
\bqn
\lb{nvectorb}
N_{\lambda}N^{\lambda} = - \frac{p_{c,T}^2}{4 N^2 p_c}.
\eqn

{A marginally trapped surface will be developed when $N_{\lambda}N^{\lambda} = 0$, or equivalently $\Theta^{+} \Theta^{-} = 0$ \cite{Hawking:1973uf,Wang:2003bt,Wang:2003xa,Gong:2007md,Hayward:1999ek}. However, as shown above, the lapse function $N$ always remains finite within a finite time \footnote{Recall that classically, we have $p^{\rm{GR}}_c(T) = e^{2T}$ and  $N^{\rm{GR}}(T) = e^{T}/\sqrt{{2m}{e^{-T}} -1}$, so that
$\Theta^{\rm{GR}}_{\pm}(T) = - e^{-T}\sqrt{2\left(2me^{-T} -1\right)}$.}. Thus, a marginally trapped surface can exist only if $p_{c, T} = 0$, which is in sharp contrast to the classical case, in which we have  $N^{\rm{GR}}\left(T^{\rm{GR}}_H\right) = \infty$ and $p_{c,T} = 2 \exp\left(2T^{\rm{GR}}_H\right)$, so that $\Theta^{\rm{GR}}_{\pm} \left(T^{\rm{GR}}_H\right)   = 0$. From the dynamical equation  (\ref{eq2.25u}), we can see that this is possible only when
\bq
\lb{MTS}
\delta_c(T) c(T) = \frac{\pi}{2}.
\eq
In Fig. \ref{fig6} we plot this quantity together with the norm, $N^{\lambda}N_{\lambda}$, from which we can clearly see that there indeed exist various points at which the above condition is satisfied, so that $p_{c,T} = 0$ at these points. In particular, in Figs. \ref{fig6} (b) and (c), we zoom in to the interval $T \in (0, 2)$, which clearly shows that two such points exist in this interval, at which $\log \left(-N^{\lambda}N_{\lambda}\right)$ becomes infinitely large, as $N^{\lambda}$ becomes null.}

{On the other hand, in  Fig. \ref{fig5} we plot several  physical quantities for $T \in (0, 2)$ including $H^{\rm{eff}}$, which shows $\left|H^{\rm{eff}}\right| \leq 2.0\times 10^{-15}$. Therefore, our numerical results are quite reliable in this interval. From this figure, we can clearly see that across these marginally trapped surfaces, all metric coefficients remain finite and non-zero.
As a result, these surfaces represent neither black hole nor white hole horizons, but transition surfaces that separate trapped regions ($\Theta^{\pm} < 0$) from the anti-trapped ones ($\Theta^{\pm} > 0$). In fact, across each of these points, $p_{c, T}$ changes its sign. Then, from Eq.(\ref{Eps}) we can see that both $\Theta_{+}$ and $\Theta_{-}$ change their signs simultaneously. Therefore, {\em these surfaces always separate trapped regions from anti-trapped ones}, while by definition a black (white) hole horizon always separates a trapped region from an untrapped one  \cite{Hawking:1973uf,Wang:2003bt,Wang:2003xa,Gong:2007md}. Therefore, we conclude that {\em in the BV model, no black/white hole structure exists}.}

{It should be noted that simply looking at Figs. \ref{m1p250T03} and \ref{m10to3p250T7}, one cannot tell the existence of these transition surfaces. This is because in these figures the quantities are plotted out in such a large range, $T \in (0, 600)$, in which the detailed changes of $p_c$ were washed out, because it has quite different values at different times. In particular, the moment
$T = 600$ corresponds to $\left.\sqrt{p_c} \right|_{T = 600} \simeq 10^{265}$ m, which is much larger than the size of our current observational universe, $L_{\rm{ob}} \simeq 8.8 \times 10^{26}$ m. }

\begin{figure*}[htbp]
\begin{tabular}{cc}
\includegraphics[width=0.45\textwidth]{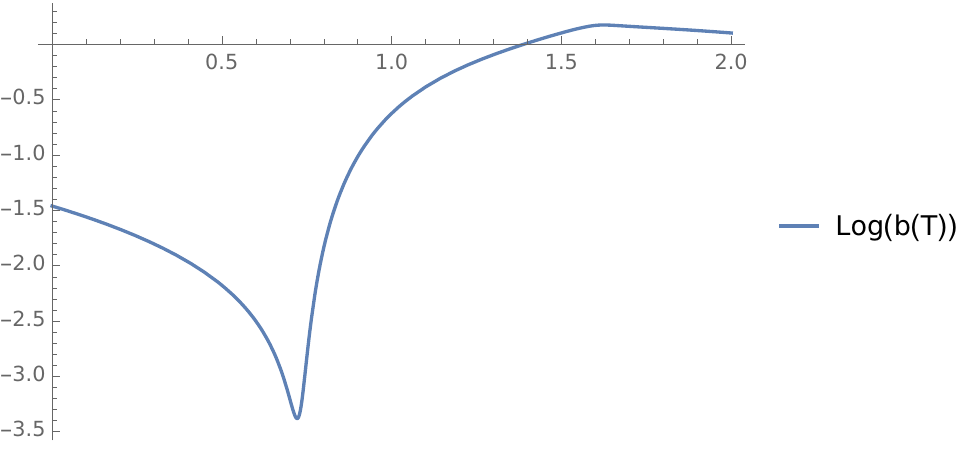}&
\includegraphics[width=0.45\textwidth]{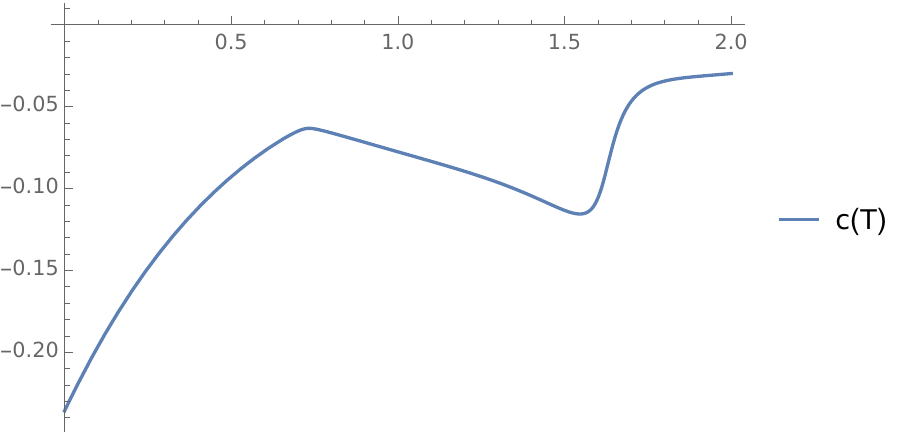}\\
\\[6pt]
\\
\includegraphics[width=0.45\textwidth]{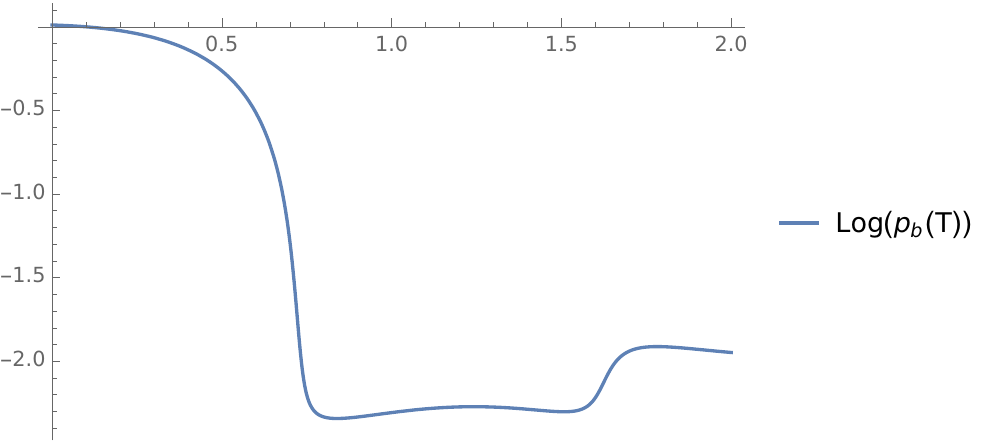}&
\includegraphics[width=0.45\textwidth]{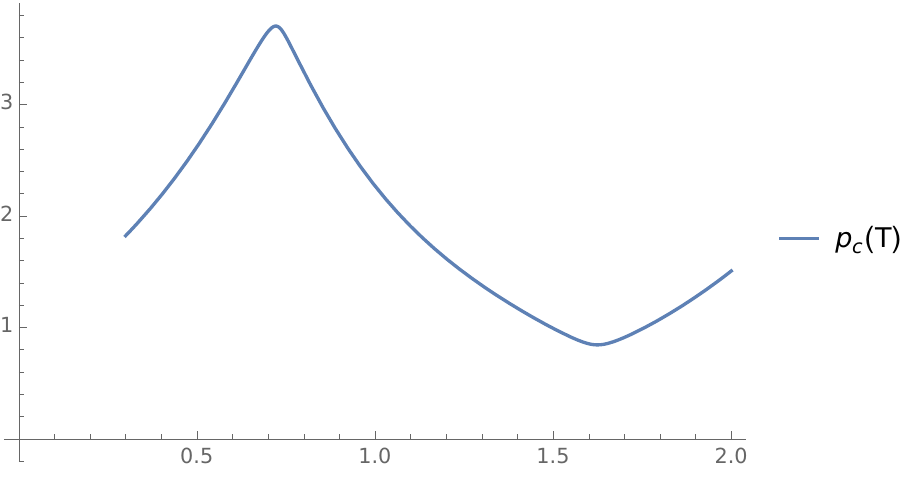}\\
\\[6pt]
\\
	\includegraphics[width=0.45\textwidth]{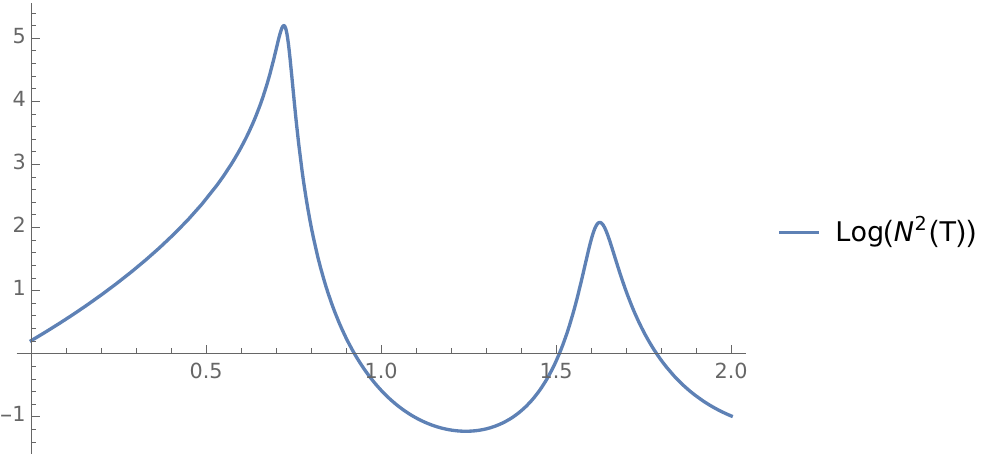}&
\includegraphics[width=0.45\textwidth]{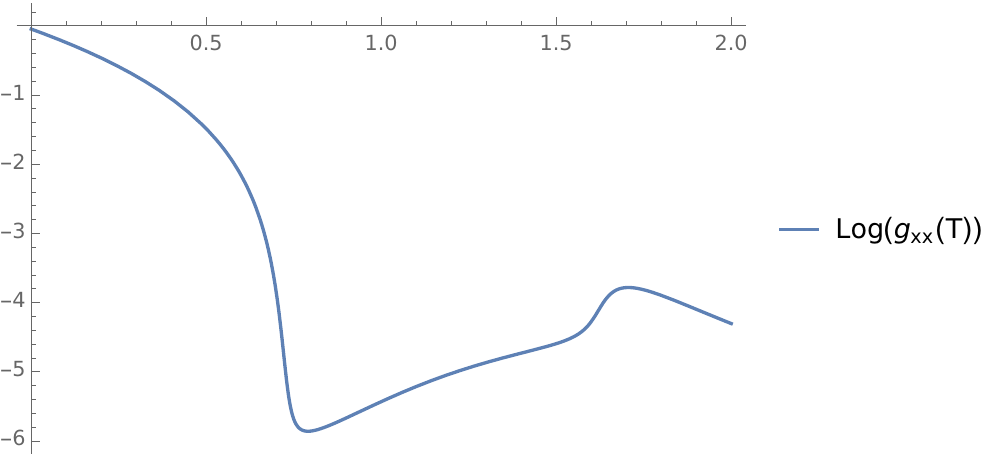}\\
\\[6pt]
\\
\includegraphics[width=0.45\textwidth]{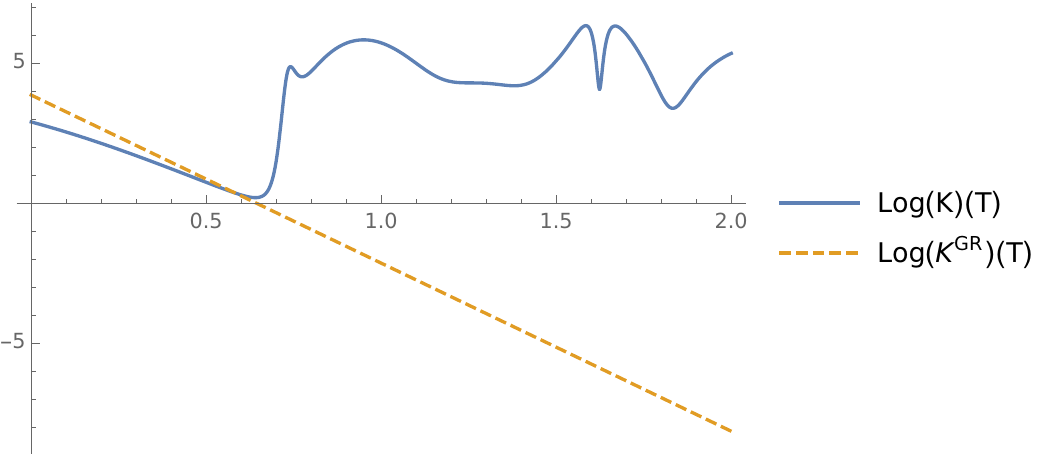}&
\includegraphics[width=0.45\textwidth]{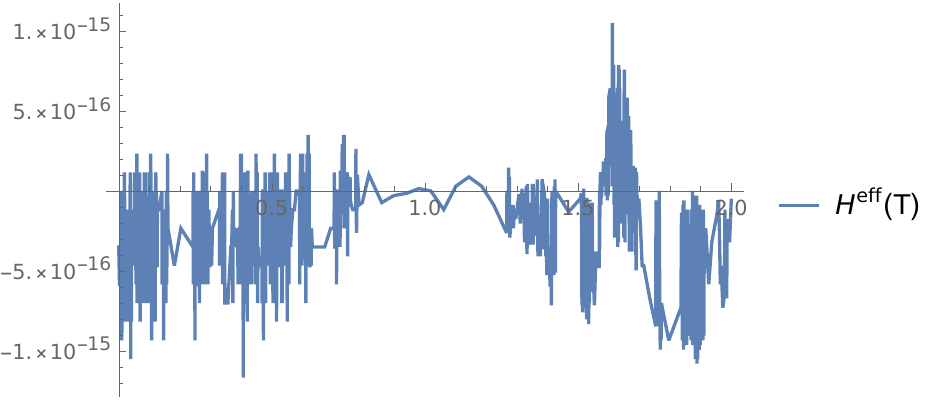}\\
\\[6pt]
	\end{tabular}
\caption{Plots of  the four physical variables $\left(b, c, p_b, p_c\right)$,  together with the metric components $N^2$, $g_{xx}$  the Kretschmann scalar $K$, and its classical counterpart $K^{\rm{GR}}(T) \equiv 48m^2/p_c^3$, and the effective Hamiltonian $H^{\rm{eff}}$ in the region $T \in (0, 2)$. The mass parameter $m$ is chosen as  $m/\ell_{pl}=1$, for which we have  $T_{\cal{T}} \simeq -1.49$ and $T^{\rm{GR}}_H \approx0.693$. The same initial time and conditions were chosen as those of Fig. \ref{m1p250T03}. In each of the subfigures, the horizontal line represents the $T$-axis.}
\lb{fig5}
\end{figure*}

\subsubsection{Asymptotic Behavior of Spacetimes as $T \ll T_{\cal{T}}$}

{To see the connection of our current studies to the ones carried out in  \cite{Boehmer:2007ket,Chiou:2008eg,Chiou:2008nm,Joe:2014tca,Dadhich:2015ora} and the global structure of the BV spacetime,  let us briefly consider the asymptotic behaviors of the spacetimes for $T \ll T_{\cal{T}}$, that is, the asymptotic behaviors of the spacetimes in the pre-transition surface. This can be shown by simply considering the case $m = \ell_{pl}$. With the same initial time and conditions as those chosen for Fig. \ref{m1p250T03}, we plot the four variables ($p_b, b; c, p_c$) in Fig. \ref{fig7}, from which we can see that they behave very much like the ones obtained in \cite{Boehmer:2007ket,Chiou:2008eg,Chiou:2008nm,Joe:2014tca,Dadhich:2015ora,Gan:2022mle}. In particular, as $T$ decreases, $p_c$ approaches a constant $\bar{p}_c$, which is smaller than the Planck area. Such a spacetime with a constant radius of the two spheres was first discussed by Nariai \cite{1999GReGr..31..945N,1951SRToh..35...46H},
and latter generalized to the charged case by Bousso \cite{Bousso:1996pn}. As shown in detail in \cite{Gan:2022mle}, in the current case, the corresponding solutions are charged Nariai solutions
\bqn
\lb{CN}
ds^2 &\simeq& \left(\frac{\bar t_0}{\bar t}\right)^2\left(- d\bar t^2 + d\bar x^2\right) + \bar{p}_c d^2\Omega\nb\\
&=& - d\hat{t}^2 + e^{2\hat{t}/\bar{t}_0}d\hat{x}^2 + \bar{p}_c d^2\Omega,
\eqn where $d\bar{t} = e^{\bar{\alpha}T}dT$, $\bar{x} = \bar{\beta} x$, and  $\bar{\alpha}, \bar{\beta}$ and $\bar{t}_0$ are all positive constants \cite{Gan:2022mle}, and $\hat{t} \equiv - \bar{t}_0 \log\bar{t}, \; \hat{x} \equiv \bar{t}_o \bar x$. From the above asymptotic behavior of the metric, we can see that it is a topology of $dS_2 \times S^2_{(0)}$, where $S^2_{(0)}$ denotes a 2-spheres with a finite radius.
As shown explicitly in \cite{Hawking:1973uf}, the coordinates ($\hat{t}, \hat{x}$) cover only part of the entire spacetime. After extension, the corresponding Penrose diagram is that given by the top part of Fig. \ref{fig8} (b)   \cite{Casals:2009zh}.}

\subsubsection{Geodesically Complete BV Spacetimes}

{It is well-known that the KS spacetime of Eq.(\ref{metric})  is usually singular in the classical theory, when filled with matter that satisfies certain energy conditions \cite{Hawking:1973uf}. The corresponding Penrose diagram is given by Fig. \ref{fig8} (a), in which each point represents a 2-sphere with the radius ${p}_c(T)$ that is a function of $T$. 
Note that the horizontal line $AB$ in Fig. \ref{fig8} (a) represents the spaceitme singularity. In the vacuum case, it corresponds to   $p_c(T = -\infty) = 0$. However, after quantum geometric effects are considered, this singularity is replaced by the transition surface ${\cal{T}}$ denoted by the curve $APB$. As shown in the last subsection, spacetime asymptotically approaches the charged Nariai solution given by Eq.(\ref{CN}) for $T \ll T_{\cal{T}}$.}

{ Combining  Figs. \ref{m1p250T03} and \ref{m10to3p250T7} with Fig. \ref{fig7}, on the other hand, we can see that the metric coefficients are all finite over the whole range $T \in (-\infty, \infty)$, which corresponds to the range $\sqrt{p_c} \in [\bar{p}_c, \infty)$, where the range of $T \in (-\infty, T_{\cal{T}})$ is mapped to $p_c \in [\bar{p}_c, p_c(T_{\cal{T}})]$, and the one of $T \in (T_{\cal{T}}, \infty)$ to  $p_c \in [p_c(T_{\cal{T}}), \infty)$. Since $\sqrt{p_c}$ represents the geometric radius of the 2-spheres of $T, x =$ Const, so $T = \infty$ (where $p_c(T = \infty) = \infty$) already represents the spacetime boundaries, represented by the line $AE$ and $BE$ in Fig. \ref{fig8} (a), and no extension beyond this point is needed.  This is also consistent with the earlier conclusions obtained in  \cite{Saini:2016vgo}. }

{On the other hand, the charged Nariai solutions given by Eq.(\ref{CN}) are also geodesically complete  \cite{Casals:2009zh}. Then, we can see that the entire BV spacetime covered by $T \in (-\infty, \infty)$ is geodesically complete. Hence, the Penrose diagram for the entire spacetime is given by Fig. \ref{fig8} (b), in which the upper quadratic part $APBDC$ corresponds to $T \in (-\infty, T_{\cal{T}})$, while the lower triangular part $AEBP$ corresponds to
$T \in (T_{\cal{T}}, \infty)$.  The two regions are smoothly connected by the transition surface  at $T = T_{\cal{T}}$ denoted by the curve $APB$. The curve $AQB$ represents the moment after which the spacertime is well approximated by the charged Nariai solutions given by Eq.(\ref{CN}).
In the past of ${\cal{T}}$,
where $T > T_{\cal{T}}$, there actually exist infinite number of transition surfaces exist, each of which separates a trapped region from an anti-trapped one.   At each transition surface, the geometric radius $\sqrt{p_c(T)}$
Is different. As $T$ increases, $\sqrt{p_c(T)}$ is getting larger and larger as shown in Figs. \ref{m1p250T03} and \ref{m10to3p250T7}. However, in the future of ${\cal{T}}$, that is, when $T < T_{\cal{T}}$, the geometric radius $\sqrt{p_c(T)}$ first gets larger and then becomes smaller and smaller, and asymptotically approaches a non-zero constant $\bar{p}_c$, as shown in Fig. \ref{fig7} (b) [See also Eq.(\ref{CN})].  }

{It must be emphasized that each point in the diagram of Fig. \ref{fig8} (b) represents 2-spheres with different radii. In particular, for $ T \ll T_{\cal{T}}$, all the 2-spheres have the same radius $\bar{p}_c$, while for $ T > T_{\cal{T}}$, the radius $\sqrt{p_c(T)}$ is time-dependent. }

\begin{figure*}[htbp]
\begin{tabular}{cc}
\includegraphics[width=0.45\textwidth]{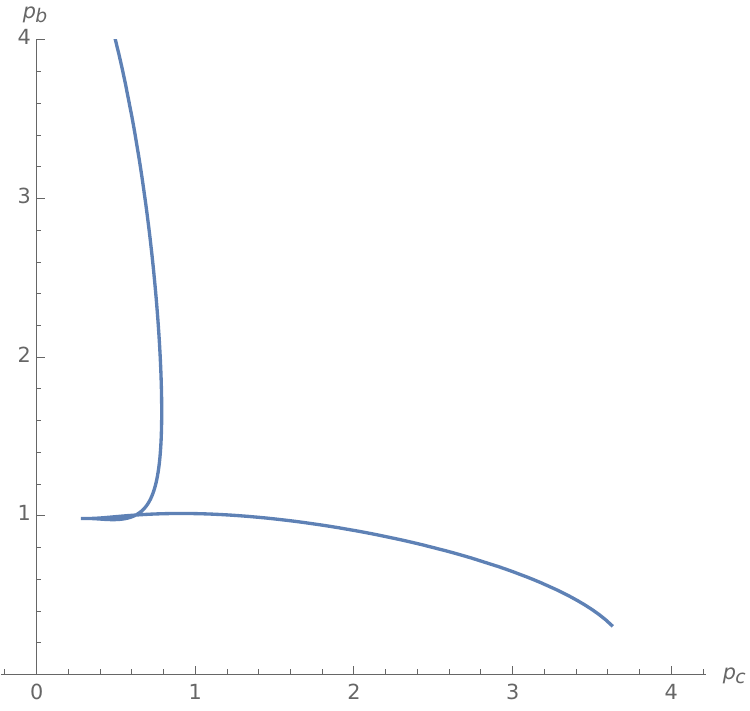}&
\includegraphics[width=0.45\textwidth]{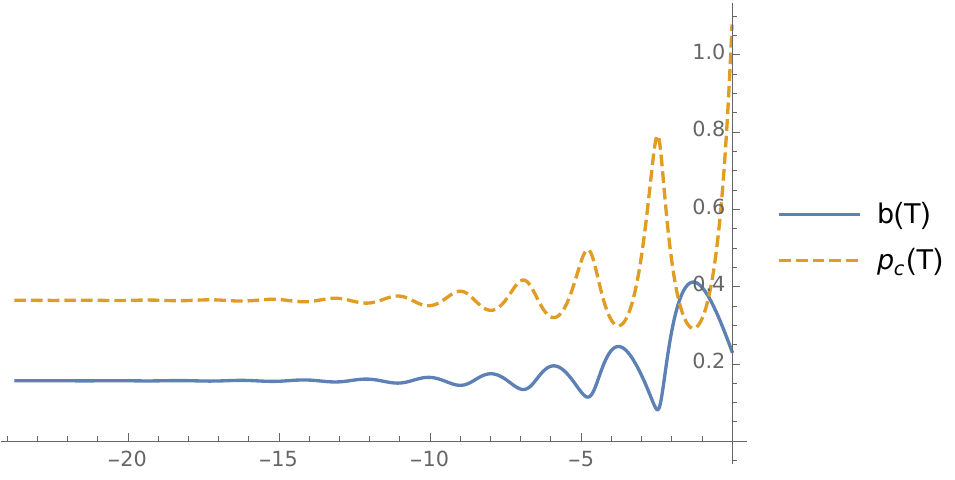}\\
(a) & (b)
\\[6pt]
\\
\includegraphics[width=0.45\textwidth]{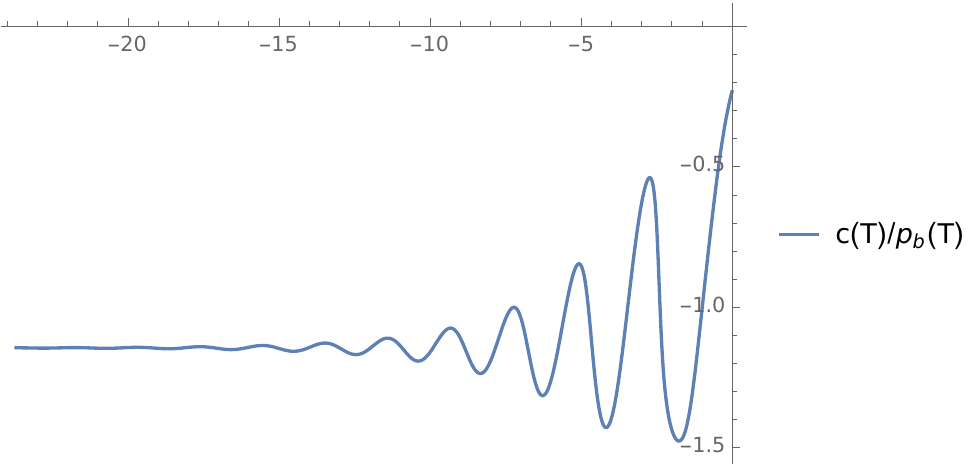}
(c)
\\[6pt]
	\end{tabular}
\caption{Plots of  the four physical variables $\left(b, c, p_b, p_c\right)$  for $T < T_{\cal{T}}$,  and  $m/\ell_{pl}=1$, for which we have  $T_{\cal{T}} \simeq -1.49$ and $T^{\rm{GR}}_H \approx0.693$.  In particular, 
 Fig. \ref{fig7} (a) is for $p_b(T)$, Fig. \ref{fig7} (b) is for $b(T)$ and $p_c(T)$, while Fig. \ref{fig7} (d) is for $c(T)/p_b(T)$.   The same initial time and conditions were chosen as those of Fig. \ref{m1p250T03}.  In each of the sub-figures, the horizontal line represents the $T$-axis.}
\lb{fig7}
\end{figure*}

\section{
Conclusions
}\lb{conclusion}
\renewcommand{\theequation}{4.\arabic{equation}}\setcounter{equation}{0}

In this paper, within the framework of the improved dynamics approach, we studied the quantum effects near the location $T  \simeq T^{\rm{GR}}_H \equiv \ln (2m)$, where the classical black hole horizon used to appear.
This was first considered by
BV \cite{Boehmer:2007ket}, in which the two polymerization parameters $\delta_b$ and $\delta_c$ are given by Eq.(\ref{eq3}), where $m$ is the classical black hole mass. To study such effects, we have chosen the initial conditions to be as close to the corresponding classical ones as possible. We found that this is always possible at a moment $T_i$, where $T_i$ satisfies the condition
$$
T_{\cal{T}} \ll T_i \ll T^{\rm{GR}}_H,
$$
see Eqs.(\ref{GRVsA}) - (\ref{Ti}) and Table \ref{ts2}, where
$T = T_{\cal{T}}$ is the location of the (first) transition surface.
To certain surprise,  we have found that a black hole (or white hole) horizon never develops within a finite time $T$ to the past of the (first) transition surface $T > T_{\cal{T}}$.
Instead, only subsequent transition surfaces exist, which always separate trapped regions from anti-trapped regions.

Combining Figs. \ref{m1p250T03}, \ref{m10to3p250T7} and \ref{fig7} with the previous studies of the BV solution in the region $T < T_{\cal{T}}$  \cite{Boehmer:2007ket,Chiou:2008eg,Chiou:2008nm}, one can see that the metric coefficients $\left(N^2, \; g_{xx}, \; p_c\right)$ and their inverses $\left(N^{-2}, \; g^{xx}, \; p_c^{-1}\right)$ are always finite and non-singular over the whole range of $T \in (-\infty, \infty)$.
Since $\sqrt{p_c}$ represents the geometric radius of the 2-spheres of $T, x =$ Const., the moment $T = \infty$, which corresponds to $p_c(T = \infty) = \infty$,
represents the spacetime boundaries, denoted by the two lines $AE$ and $BE$ in   Fig. \ref{fig8} (a), and no extension of the spacetime beyond them is needed. Then, the spacetime covered by $T \in (-\infty, \infty)$  is geodesically complete, which is consistent with the analytical conclusions obtained in  \cite{Saini:2016vgo}.

In such a geodesically complete spacetime, we have shown that a black/white hole horizon is never developed. This can be understood as follows: In the choice of the two parameters $\delta_b$ and $\delta_c$, BV imposed the condition
\bq
\lb{eq4.1}
\delta_c p_b = \sqrt{\Delta p_c}.
\eq
In the classical case, $p_b$ vanishes at the black hole horizon $T = T^{\rm{GR}}_H$, while $p_c$ remains finite, Eq.(\ref{eq4.1}) tells us that   $\delta_c \rightarrow \infty$, which in turn implies that the quantum effects become infinitely large [cf. Eq.(\ref{bv-delta_c})]. As a result, both $\delta_c$ and   $p_b$ will remain finite and non-zero over the whole spacetime, and a black/white hole horizon is never developed \footnote{With similar  arguments, we would like to point out that near the classical spacetime singularity $p_c^{\rm{GR}} \simeq 0$, the parameter $\delta_b$ becomes very large, as can be seen from Eq.(\ref{deltab}), i.e.
$\delta_b = \sqrt{{\Delta}/{p_c}}$.
Then, following the arguments given above, we can see that quantum effects become extremely large near the classical singularity, and as a result, the classical singularity is finally resolved and replaced by a quantum bounce.}. This observation is important, and one should pay attention to its usage whenever the $\bar{\mu}$-scheme is applied  to LQBHs \cite{Olmedo:2016ddn,Ashtekar:2018cay,Ashtekar:2020ifw,Gambini:2022hxr,Ashtekar:2023cod}.

The above-mentioned properties are sharply contrasted to those obtained in other models of LQBHs studied so far \cite{Olmedo:2016ddn,Ashtekar:2018cay,Ashtekar:2020ifw,Gambini:2022hxr,Ashtekar:2023cod}, and put  the BV model as physically viable is more questionable
\cite{Boehmer:2007ket,Corichi:2015xia,Olmedo:2017lvt,Ashtekar:2018cay}.

It must be noted that this by no means implies that the improved dynamics approach is not applicable to the loop quantization of black holes, but rather that the starting point should not be the KS spacetime.  In fact, in \cite{Han:2020uhb} Han and Liu started with the Lemaitre-Tolman-Bondi (LTB) metric
\bqn
\lb{eq4.3}
ds^2 = - dt^2 + \frac{\left(E^{\varphi}\right)^2}{|E^x|} dx^2 + |E^x| d^2\Omega,
\eqn and found that the quantum effects are negligible near a macroscopic black hole horizon using the same improved dynamics approach as adopted by BV. This is precisely because the physical distance along the $x$-direction in the LTB coordinates now becomes finite near the black hole horizon \cite{Zhang:2023noj}. In fact, the Schwarzschild black hole solution in LTB coordinates is given by
\bqn
\lb{eq4.4}
E^x_{\text{GR}} &=& \left[\frac{3}{2}\sqrt{2m}(x - t)\right]^{4/3}, \nb\\
E^{\varphi}_{\text{GR}} &=& \frac{2}{3} \left(\frac{3}{2}\sqrt{2m}\right)^{4/3}(x - t)^{1/3}.
\eqn
The spacetime singularity is now located at $x - t = 0$, while the black hole horizon is located at $x - t =4m/3$. Denoting the conjugate momentum of $E^x$ and $E^{\varphi}$ by $K_x$ and $K_{\varphi}$, respectively, Han and Liu considered the following replacements \cite{Han:2020uhb}
\bqn
\lb{eq3.3}
K_x \rightarrow \frac{\sin(\delta_x K_x)}{\delta_x}, \quad K_{\varphi} \rightarrow \frac{\sin(\delta_{\varphi} K_{\varphi})}{\delta_{\varphi}}.
\eqn where
\bqn
\lb{eq3.4}
\delta_x = \frac{2\gamma \sqrt{\Delta |E^x|}}{E^{\varphi}}, \quad  \delta_{\varphi} = \frac{\gamma \sqrt{\Delta}}{\sqrt{|E^x|}}.
\eqn
Clearly, near the singularity we have  $(\delta_x, \delta_{\varphi}) \rightarrow (0, \infty)$. Then, it is expected that quantum gravitational effects will become very large; therefore, in reality, the singularity that appeared classically is now smoothed out by these quantum effects, and a non-singular transition surface finally replaces the singularity. On the other hand, near the location of the classical black hole horizon, we have $(\delta_x, \delta_{\varphi})  \simeq \gamma \sqrt{\Delta}(2, (2m)^{-1})$, which are all finite. These conclusions are consistent with the results obtained in \cite{Han:2020uhb}.

It should also be noted that BV considered only the case where $\delta_{b}$ and $\delta_{c}$ are functions of $p_b$ and $p_c$. In general, they can be functions of the four variables, $b, c, p_b$ and $p_c$. Therefore, it would be very interesting to see how our above conclusions are affected by such dependence.

Finally, we note that the formation of LQBHs from gravitational collapse has been studied extensively in the last couple of years; see, for example, \cite{Giesel:2023hys,Fazzini:2023ova,Lewandowski:2022zce}, and references therein. It is very important to understand how the LQBHs found so far can be formed in our Universe from the gravitational collapse of realistic matter.

\Acknowledgements{
We would like to thank Prof. Parampreet Singh for his valuable discussions and comments. Our thanks also go to Drs. Jared Fier and John Vasut for reading our manuscript carefully and providing valuable suggestions and comments. Wen-cong Gan was supported by Baylor University through the Baylor Physics graduate program and also by the Initial Research Foundation of Jiangxi Normal University under the Grant No. 12022827.
Numerical computations were performed on the public computing service platform provided by TianHe-2 through the Institute for Theoretical Physics \& Cosmology, Zhejiang University of Technology.
This work was partially supported by the National Key Research and Development Program of China under Grant No. 2020YFC2201504, the National Natural Science Foundation of China with Grants No. 11975203, No. 12075202, and No.11875136, and the Natural Science Foundation of Jiangsu Province under Grant No. BK20211601.
Anzhong Wang is partially supported by a US NSF grant (Grant number: PHY2308845).}

\bibliographystyle{scichina}
\bibliography{BV}

\end{multicols}

\end{document}